\documentclass[10pt]{iopart}

\usepackage{iopams}
\usepackage{braket}
\usepackage{cite}
\usepackage{color}
\usepackage{graphicx}
\usepackage{ulem}
\usepackage{siunitx}

\begin{document}
\date{\today}
\topical[]{Approaches to tunnel magnetoresistance effect with antiferromagnets}
\author{Katsuhiro Tanaka}
\address{Department of Physics, University of Tokyo, Hongo, Bunkyo-ku, Tokyo 113-0033, Japan}
\author{Takuya Nomoto}
\address{Department of Physics, Tokyo Metropolitan University, Hachioji, Tokyo 192-0397, Japan}
\author{Ryotaro Arita}
\address{Department of Physics, University of Tokyo, Hongo, Bunkyo-ku, Tokyo 113-0033, Japan}
\address{Center for Emergent Matter Science, RIKEN, Wako, Saitama 351-0198, Japan}
\vspace{10pt}
\begin{indented}
	\item[]\today
\end{indented}
\begin{abstract}
The tunnel magnetoresistance (TMR) effect is one of the representative phenomena in spintronics.
Ferromagnets, which have a net spin polarization, have been utilized for the TMR effect.
Recently, by contrast, the TMR effect with antiferromagnets, which do not possess a macroscopic spin polarization, has been proposed, 
and also been observed in experiments.
In this topical review, we discuss recent developments in the TMR effect, particularly focusing on the TMR effect with antiferromagnets.
First, we review how the TMR effect can occur in antiferromagnetic tunnel junctions.
The Julliere model, which has been conventionally utilized to grasp the TMR effect with ferromagnets,
breaks down for the antiferromagnetic TMR effect.
Instead, we see that the momentum dependent spin splitting explains the antiferromagnetic TMR effect.
After that, we revisit the TMR effect from viewpoint of the local density of states (LDOS).
We particularly focus on the LDOS inside the barrier,
and show that the product of the LDOS will qualitatively capture the TMR effect not only in the ferromagnetic tunnel junctions but also in the ferrimagnetic and antiferromagnetic tunnel junctions.
This method is expected to work usefully for designing magnetic tunnel junctions.
\end{abstract}
\maketitle
%
%
%
%
\ioptwocol
%
\section{Introduction}
\label{sec:introduction}
The tunnel magnetoresistance (TMR) effect~\cite{Julliere1975_PhysLettA_54A_225} is one of the typical phenomena in the field of spintronics~\cite{Zutic2004_RevModPhys_76_323,Felser2007_AngewChemIntEd_46_668,Chappert2007_NatMater_6_813,Bader2010_AnnuRevCondensMatterPhys_1_71,Hirohata2020_JMagnMagnMater_509_166711}.
The TMR effect is observed in multilayer systems, magnetic tunnel junctions (MTJs), 
which have two magnetic electrodes sandwiching an insulating thin film.
By applying a voltage, the tunneling current can flow through the MTJ as a quantum mechanical effect.
Depending on the relative directions of the magnetic moments between two magnetic electrodes,
the tunneling resistance can change.
In addition to the physics viewpoint, 
namely, the ballistic transport properties of electrons in relation to the spin degrees of freedom,
the TMR effect has attracted attention from the application point of view.
The MTJ is utilized for the devices such as the magnetic random access memories (MRAM) or the magnetic head of the hard-disk drive.
\par
After the discovery of the TMR effect, 
the TMR effect has been discussed using ferromagnets which has a macroscopic spin polarization~\cite{Julliere1975_PhysLettA_54A_225,Maekawa1982_IEEETranMagn_18_707,Slonczewski1989_PhysRevB_39_6995,Miyazaki1995_JMagnMagnMater_139_L231,Moodera1995_PhysRevLett_74_3273,Mathon1997_PhysRevB_56_11810,Butler2001_PhysRevB_63_054416,Mathon2001_PhysRevB_63_220403,Parkin2004_NatMater_3_862,Yuasa2004_NatMater_3_868,Djayaprawira2005_ApplPhysLett_86_092502,Ikeda2008_ApplPhysLett_93_082508,Ikeda2010_NatMater_9_721,Worledge2011_ApplPhysLett_98_022501,Scheike2023_ApplPhysLett_122_112404,Tsymbal2003_JPhysCondensMatter_15_R109,Zhang2003_JPhysCondensMatter_15_R1603,Ito2006_JMagnSocJpn_30_1,Yuasa2007_JPhysD_40_R337,Butler2008_SciTechnolAdvMater_9_014106}.
There, it has been believed that the magnetic electrode with a large spin polarization on the Fermi level is important to generate a large TMR ratio.
\par
Recently, on the other hand, 
the TMR effect has taken renewed interest with antiferromagnets which do not have a net spin polarization;
recent developments have proposed that a particular kind of antiferromagnets can be utilized for the spintronic phenomena, namely, they can be controlled by an external field and can be incorporated to the spintronic devices even without net magnetization~\cite{MacDonald2011_PhilTransRSocA_369_3098,Gomonay2014_LowTempPhys_40_17,Jungwirth2016_NatNanotechnol_11_231,Baltz2018_RevModPhys_90_015005,Zelezny2018_NatPhys_14_220,Siddiqui2020_JApplPhys_128_040904,Amin2020_ApplPhysLett_117_010501,Fukami2020_JApplPhys_128_070401,Chen2024_AdvMater_36_2310379}.
The TMR effect using antiferromagnets has been actually proposed,
and a finite TMR effect has been observed in the all antiferromagnetic tunnel junctions based on $\mathrm{Mn_{3}Sn}$ or $\mathrm{Mn_{3}Pt}$~\cite{Chen2023_Nature_613_490,Qin2023_Nature_613_485,Shi2024_AdvMater_36_2312008}.
\par
This topical review aims to understand such antiferromagnetic TMR effect, 
and explore a way toward designing the MTJs.
For the ferromagnetic MTJs, the conventional Julliere model has still somewhat survived,
although a more microscopic mechanism, namely, coherent tunneling, 
has been discussed for more closer understanding.
However, in the antiferromagnetic MTJs, the Julliere model completely breaks down since antiferromagnets do not have a macroscopic spin polarization which needs to generate the TMR effect in the model.
Therefore, we need another picture to simply describe the antiferromagnetic TMR effect.
In this review, we will introduce the role of momentum-dependent spin polarization of antiferromagnets for the TMR effect;
we will show that the antiferromagnetic TMR effect can be described by a finite spin splitting in the momentum space which antiferromagnets breaking the time-reversal symmetry have.
\par
The proposal of the TMR effect with antiferromagnets will increase the possibility of utilizing a wider variety of magnetic materials and also insulating barriers compatible with those magnets.
Given this situation, it is highly desired to develop a method which can efficiently evaluate the TMR property of the antiferromagnetic MTJs as well as the ferromagnetic MTJs for further exploration of the MTJs.
In this topical review, we next focus on the local density of states (LDOS) in the MTJs.
We will show that the product of the LDOS of the barrier region can trace the TMR effect qualitatively,
which can also be understood as a simple extension of the Julliere model incorporating the tunneling effect and the details of the materials.
\par
The remainder of this review is as follows;
after briefly reviewing the ferromagnetic TMR effect first,
we discuss the recent developments of the TMR effect with antiferromagnetic electrodes ~\cite{Chen2023_Nature_613_490,Merodio2014_ApplPhysLett_105_122403,Stamenova2017_PhysRevB_95_060403,Zelezny2017_PhysRevLett_119_187204,Jia2020_SciChinnaPhys_63_297512,Shao2021_NatCommun_12_7061,Smejkal2022_PhysRevX_12_011028,Dong2022_PhysRevLett_128_197201,Qin2023_Nature_613_485,Tanaka2023_PhysRevB_107_214442,Cui2023_PhysRevB_108_024410,Gurung2023_arXiv_2306.03026,Jia2023_PhysRevB_108_104406,Jiang2023_PhysRevB_108_174439,Chi2024_PhysRevApplied_21_034038,Samanta2024_PhysRevB_109_174407,Shi2024_AdvMater_36_2312008,Zhu2024_JMagnMagnMater_597_172036,Zhu2024_ChinPhysLett_41_047502,Shao2024_npjSpintronics_2_13}.
We see that the spin splitting in the momentum space works for generating a finite TMR effect in the antiferromagnetic electrodes,
and refer to the theoretical and experimental studies of the antiferromagnetic TMR effect.
In the latter part, we discuss a real-space approach to the TMR effect using the LDOS inside the insulating barrier based on Ref.~\cite{Tanaka2023_PhysRevB_107_214442}.
After confirming that the LDOS works as an indicator for the TMR effect with the lattice models,
we present that the LDOS will be useful for designing the MTJs using the ferromagnetic Fe/MgO/Fe and the antiferromagnetic $\mathrm{Ru}_{1-x}\mathrm{Cr}_{x}\mathrm{O}_{2}/\mathrm{TiO_{2}}/\mathrm{Ru}_{1-x}\mathrm{Cr}_{x}\mathrm{O}_{2}$ tunnel junctions as concrete examples.
\section{Tunnel magnetoresistance effect with antiferromagnets: role of momentum dependent spin splitting}
\label{sec:afm_tmr}
\subsection{Tunnel magnetoresistance effect with ferromagnetic electrodes}
\label{subsec:afm_tmr_ferro}
\begin{figure}[tbh]
	\centering
	\includegraphics[width=80mm]{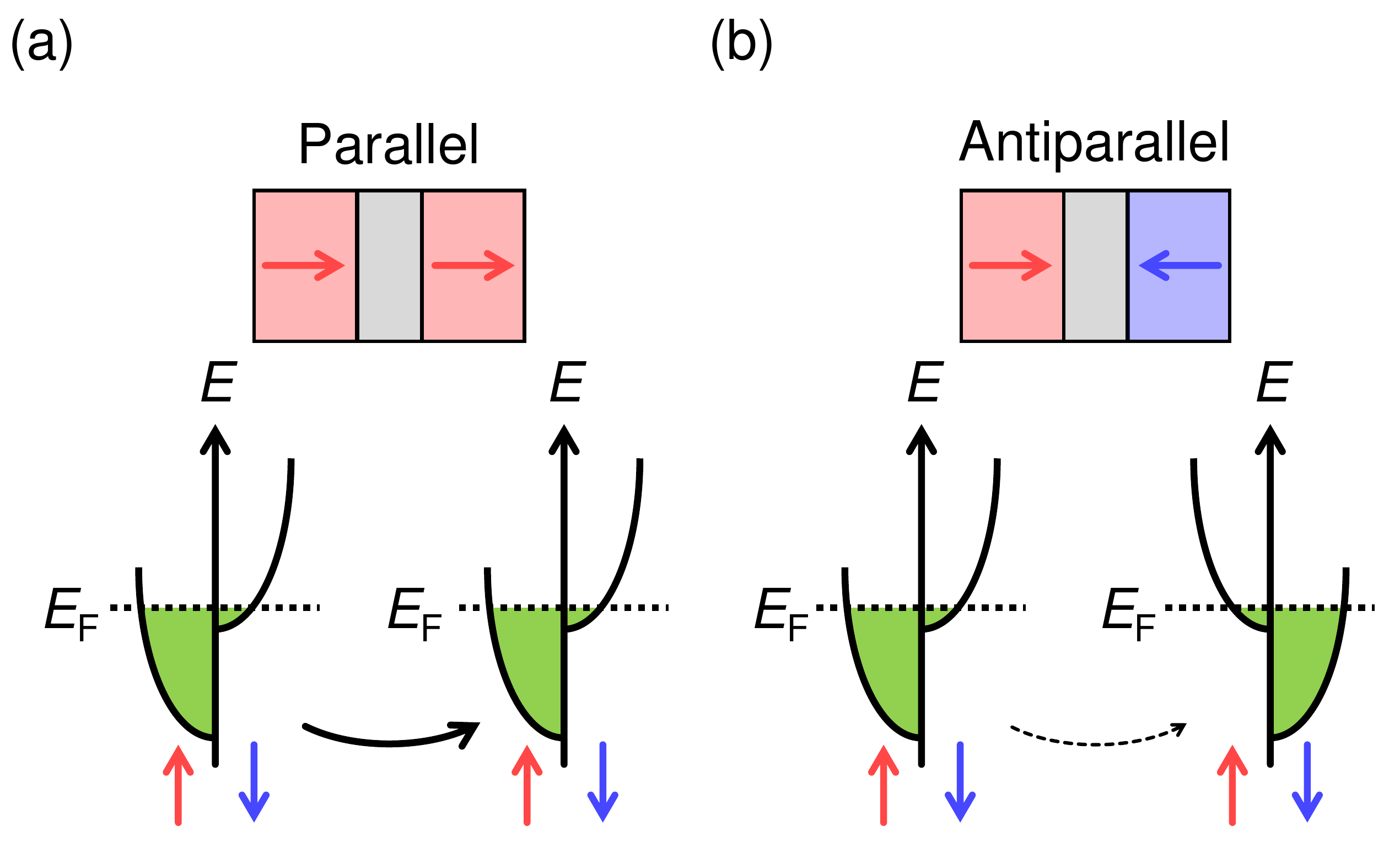}
	\caption{%
		Schematic view of tunnel magnetoresistance effect with ferromagnets for (a) the parallel and (b) antiparallel configurations from the Julliere model.
		Upper panels represent the magnetic tunnel junctions. Red and blue rectangles show the magnetic electrodes, and gray rectangles show the insulating barrier.
		Arrows show the directions of the magnetic moment in each electrode.
		Lower panels represent the spin-resolved density of states of the magnetic electrodes.
	}
	\label{fig:julliere_tmr}
\end{figure}
First, we review the conventional TMR effect with ferromagnets briefly.
The simplest model to grasp the ferromagnetic TMR effect is the so-called Julliere model~\cite{Julliere1975_PhysLettA_54A_225}.
The Jullire model explains the TMR effect using the macroscopic spin polarization of the magnetic electrodes on the Fermi level.
Namely, in the Julliere model,
the tunneling conductance of the MTJ, $\tau_{\text{DOS}}(E)$, 
is estimated as
\begin{eqnarray}
	\tau_{\text{DOS}}(E)
= \sum_{\sigma} D_{\text{L}, \sigma}(E) D_{\text{R}, \sigma}(E).
\label{eq:dos_product}
\end{eqnarray}
Here, $D_{\text{L/R}, \sigma}(E)$ ($\sigma = \uparrow, \downarrow$) is the bulk density of states (DOS) with spin-$\sigma$ of the left/right magnetic electrodes.
For simplicity, let us assume that the left and right electrodes are the same material,
and $D_{\uparrow}(E)/(D_{\uparrow}(E)+D_{\downarrow}(E)) = x$.
Then, $\tau_{\text{DOS}}(E)$ for the parallel/antiparallel configuration, $\tau_{\text{DOS, P/AP}}(E)$, are written as
\begin{eqnarray}
	\tau_{\text{DOS, P}}(E) 
& = (x^{2} + (1-x)^{2})(D_{\uparrow}(E)+D_{\downarrow}(E))^{2}, \\
	\tau_{\text{DOS, AP}}(E) 
& = 2x(1-x) (D_{\uparrow}(E)+D_{\downarrow}(E))^{2},
\end{eqnarray}
and $\tau_{\text{DOS, P}}(E) - \tau_{\text{DOS, AP}}(E) \propto (2x-1)^{2}$ holds.
When we consider the ferromagnetic electrodes,
we have the spin polarization at the Fermi energy, namely, $x \neq 0.5$,
and thus $\tau_{\text{DOS, P}}(E) > \tau_{\text{DOS, AP}}(E)$ holds.
\par
When the Fermi level is perfectly spin polarized, namely, $x = 0$ or $1$, 
$\tau_{\text{DOS, P}}(E) - \tau_{\text{DOS, AP}}(E)$ takes the largest value.
Namely, the materials with half-metallic electronic states will be useful for the realization of a large TMR ratio. 
From this viewpoint, half-Heusler metals have been explored as a promising candidate for realizing the half-metallic TMR effect~\cite{Sakuraba2005_JpnJApplPhys_44_L1100,Sakuraba2006_ApplPhysLett_88_192508}.
\par
In reality, however, one should consider the details of the MTJ such as the modulation of the electronic structure at the interface between the electrodes and the barrier, or the decaying properties inside the insulators, 
to precisely understand the TMR effect,
while the above Julliere model ignores such details.
Butler~\textit{et al.} gave a microscopic view~\cite{Butler2001_PhysRevB_63_054416} using the bcc-Fe/MgO/Fe MTJ as a specific example.
They focused on the symmetries of the tunneling electrons.
The electronic states around the Fermi level of bcc-Fe consist of the electrons with the $\Delta_{1}$-symmetry, whose electron orbitals are rotationally symmetric around $z$-axis.
In addition, in the MgO barrier, this $\Delta_{1}$ state has a lower decay rate than the states with other symmetries like $\Delta_{2}$ states or $\Delta_{5}$ states.
Therefore, only the spin-polarized $\Delta_{1}$ state can tunnel through the MgO barrier,
which will lead to a large TMR ratio.
Utilizing this strategy, Yuasa~\textit{et al.} observed a large TMR ratio at room temperature in the Fe/MgO/Fe epitaxial tunnel junction~\cite{Yuasa2004_NatMater_3_868}; 
the epitaxial MTJ made it possible for electrons to tunnel through the MTJ keeping the orbital symmetry.
\subsection{Tunnel magnetoresistance effect with antiferromagnetic electrodes}
\label{subsec:afm_tmr_antiferro}
\begin{figure}[tbh]
	\centering
	\includegraphics[width=80mm]{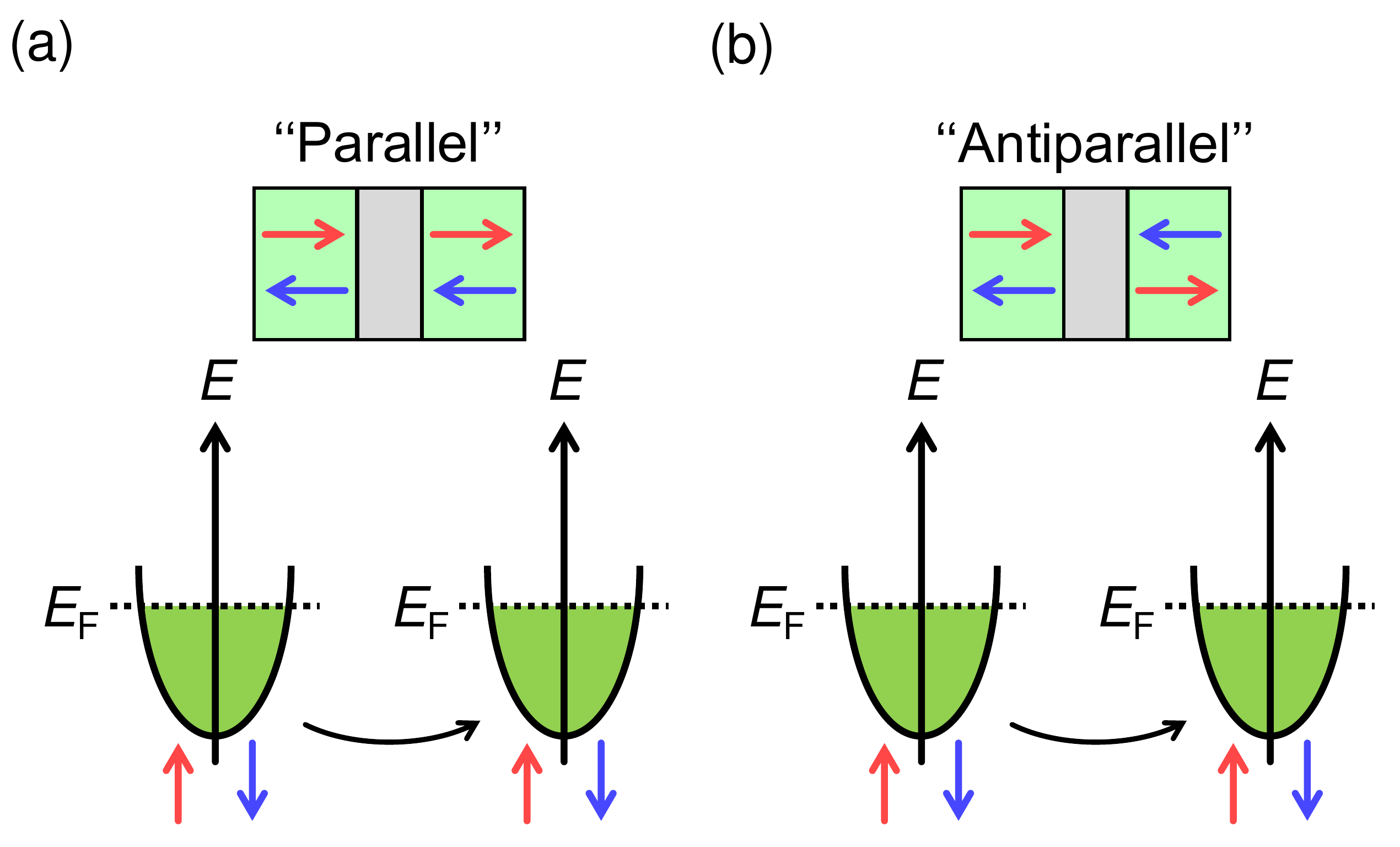}
	\caption{%
		Schematic picture of antiferromagnetic tunnel magnetoresistance effect viewed from the Julliere model, which gives the same tunneling conductance for the (a) ``parallel'' and (b) ``antiparallel'' configurations.
		Upper panels show the magnetic configurations of the magnetic tunnel junctions.
		Lower panels represent the spin-resolved density of states of antiferromagnetic electrodes.
	}
	\label{fig:julliere_tmr_antiferro}
\end{figure}
Next, we consider the case where the magnetic electrodes are antiferromagnetic materials.
When we naively see this in terms of the Julliere's picture,
the total spin polarization, namely, the DOS of the majority and minority spins, are the same,
as schematically shown in Fig.~\ref{fig:julliere_tmr_antiferro},
and the TMR effect should be zero.
We can understand this situation also by considering the $x = 0.5$ case in the discussion in Sec.~\ref{subsec:afm_tmr_ferro};
$\tau_{\text{DOS, P}}(E) - \tau_{\text{DOS, AP}}(E) = 0$ holds.
Also, we should be careful how we can define the parallel and antiparallel configurations.
\par
Recently, however, it has been proposed that one can realize the TMR effect even with antiferromagnets, 
by examining the structures of the MTJs and the characteristics of antiferromagnets more closely.
Here we introduce the antiferromagnetic TMR effect for two cases;
one is the TMR effect owing to the interfacial structure of the MTJ,
and the other exhibits a finite TMR effect originating from the magnetic and electronic properties of antiferromagnets breaking the time-reversal symmetry.
\subsubsection{Antiferromagnet with controlled interface}
\label{subsubsec:afm_tmr_antiferro_interface}
\begin{figure}[tbh]
	\centering
	\includegraphics[width=80mm]{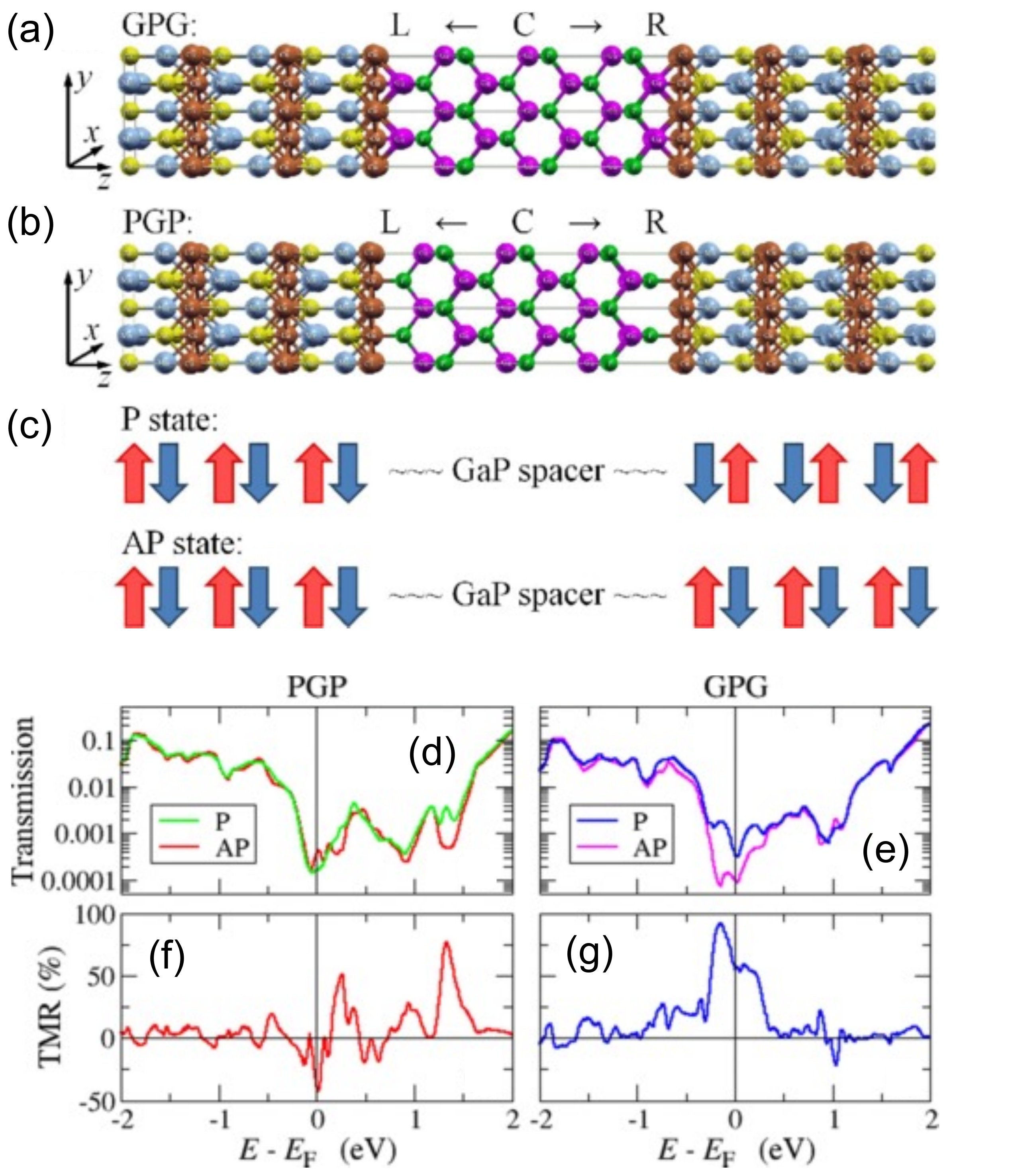}
	\caption{%
		First-principles calculation of the tunnel magnetoresistance effect in the $\mathrm{CuMnAs}$/$\mathrm{GaP}$/$\mathrm{CuMnAs}$ magnetic tunnel junction (MTJ).
		(a), (b) Crystal structures of the $\mathrm{CuMnAs}$/$\mathrm{GaP}$/$\mathrm{CuMnAs}$ MTJ.
		Termination of the $\mathrm{GaP}$ barrier is (a) $\mathrm{Ga}$-layer and (b) $\mathrm{P}$-layer.
		(c) Schematics of the parallel and antiparallel configurations of the $\mathrm{CuMnAs}$/$\mathrm{GaP}$/$\mathrm{CuMnAs}$ MTJ.
		(d)--(g) Transmission properties in the $\mathrm{CuMnAs}$/$\mathrm{GaP}$/$\mathrm{CuMnAs}$ MTJ.
		(d), (e) Energy dependence of the transmissions for the MTJ with (d) $\mathrm{P}$-termination and (e) $\mathrm{Ga}$-termination.
		(f), (g) Energy dependence of the TMR ratio calculated by $(T_{\text{P}}-T_{\text{AP}})/(T_{\text{P}}+T_{\text{AP}})$ for the MTJ with (f) $\mathrm{P}$-termination and (g) $\mathrm{Ga}$-termination,
		where $T_{\text{P}}$ and $T_{\text{AP}}$ are the transmissions for the parallel and antiparallel configurations, respectively.
		Figures are adopted from Ref.~\cite{Stamenova2017_PhysRevB_95_060403} (\copyright American Physical Society (2017)).
	}
	\label{fig:stamenova}
\end{figure}
First, we introduce the antiferromagnetic TMR effect in the MTJ whose interfacial structure is well controlled.
Stamenova~\textit{et al.} studied the MTJ with a collinear antiferromagnet $\mathrm{CuMnAs}$ with tetragonal crystal structure~\cite{Stamenova2017_PhysRevB_95_060403},
which is one of the promising candidate materials utilized for antiferromagnetic spintronics~\cite{Wadley2016_Science_351_587,Olejnik2017_NatCommun_8_15434,Godinho2018_NatCommun_9_4686}.
They considered the $\mathrm{CuMnAs}$/$\mathrm{GaP}$/$\mathrm{CuMnAs}$ tunnel junction.
They examined multiple sets of configurations depending on the interfacial structures,
namely, $\mathrm{Ga}$-layers of the $\mathrm{GaP}$-barrier face to the magnetic electrodes (GPG) (Fig.~\ref{fig:stamenova}(a)),
or $\mathrm{P}$-layers face to the magnetic electrodes (PGP) (Fig.~\ref{fig:stamenova}(b)).
They calculated the transmission for each configuration (Figs.~\ref{fig:stamenova}(d)~and~\ref{fig:stamenova}(e)),
and showed that the $\mathrm{CuMnAs}$-based MTJ exhibits a finite TMR effect (Figs.~\ref{fig:stamenova}(f)~and~\ref{fig:stamenova}(g)).
Such type of the MTJs with multiple sets of the interfacial structure has also been discussed with an antiferromagnet $\mathrm{Mn_{2}Au}$~\cite{Jia2020_SciChinnaPhys_63_297512,Jia2023_PhysRevB_108_104406,Zhu2024_JMagnMagnMater_597_172036} as well as the lattice model calculation~\cite{Saidaoui2017_PhysRevB_95_134424}.
\par
It should be noted that one needs to precisely control the interface to realize the situation when we utilize the antiferromagnets shown above.
Otherwise, the TMR properties will be cancelled out, and the TMR ratio will become almost zero in total.
Such cancellation has been found in a ferrimagnet/insulator/ferromagnet tunnel junction~\cite{Jeong2016_NatCommun_7_10276};
the observed TMR ratio is small owing to the multiple kinds of interfaces between the ferrimagnet and the insulator, 
while a large TMR ratio using the ferrimagnet has been predicted by first-principles calculation~\cite{Miura2014_IEEETransMag_50_1400504}.
\subsubsection{Antiferromagnet breaking time-reversal symmetry}
\label{subsubsec:afm_tmr_antiferro_breakingtrs}
\begin{figure}[tbh]
	\centering
	\includegraphics[width=80mm]{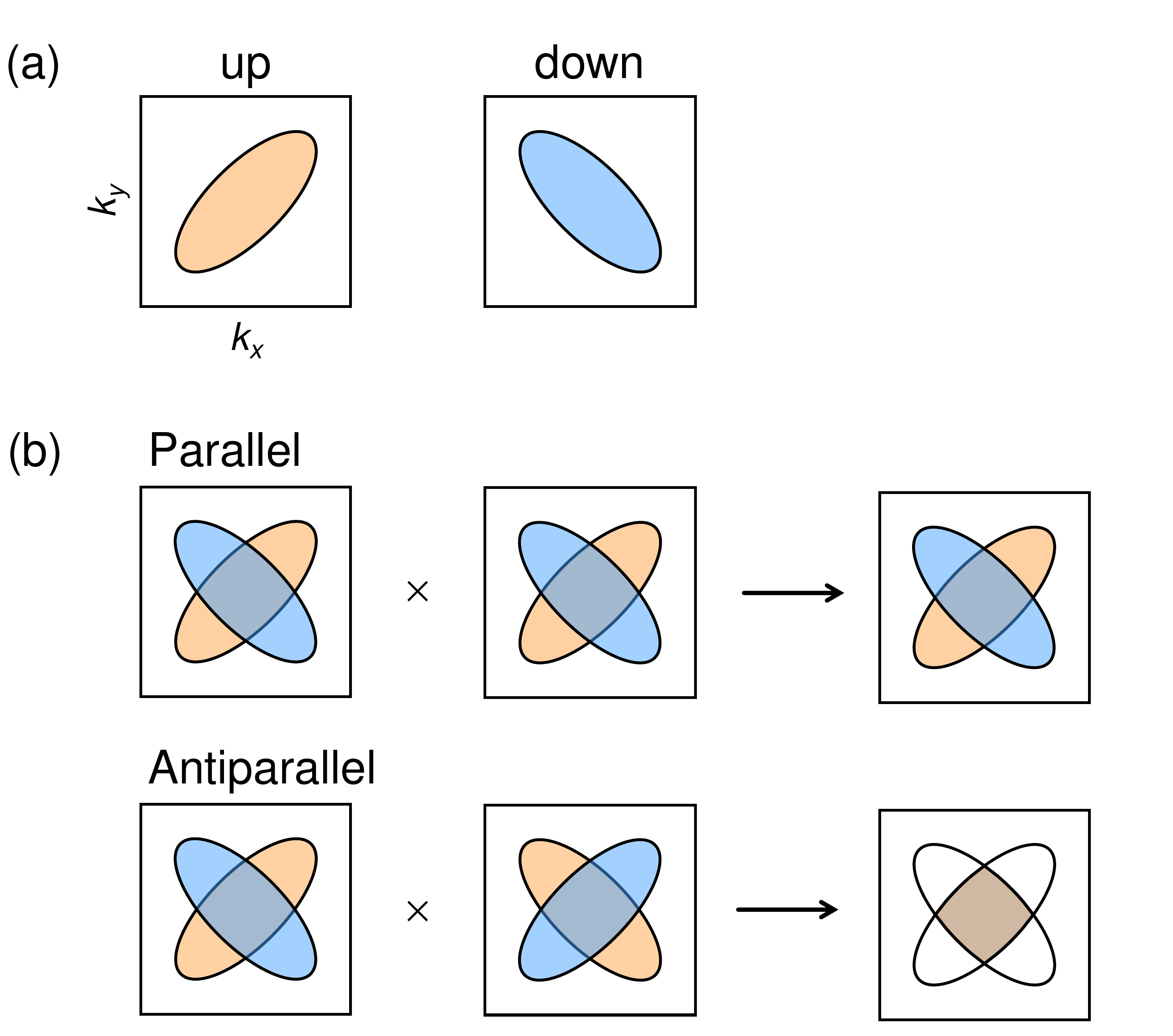}
	\caption{%
		Schematics of tunnel magnetoresistance effect with antiferromagnets breaking time-reversal symmetry.
		(a) Schematic view of the spin splitting in the momentum space at the Fermi level for antiferromagnets breaking time-reversal symmetry.
		(b) Schematic view of the antiferromagnetic tunnel magnetoresistance effect for parallel and antiparallel configurations.
	}
	\label{fig:afm_tmr}
\end{figure}
\begin{figure*}[tbh]
	\centering
	\includegraphics[width=160mm]{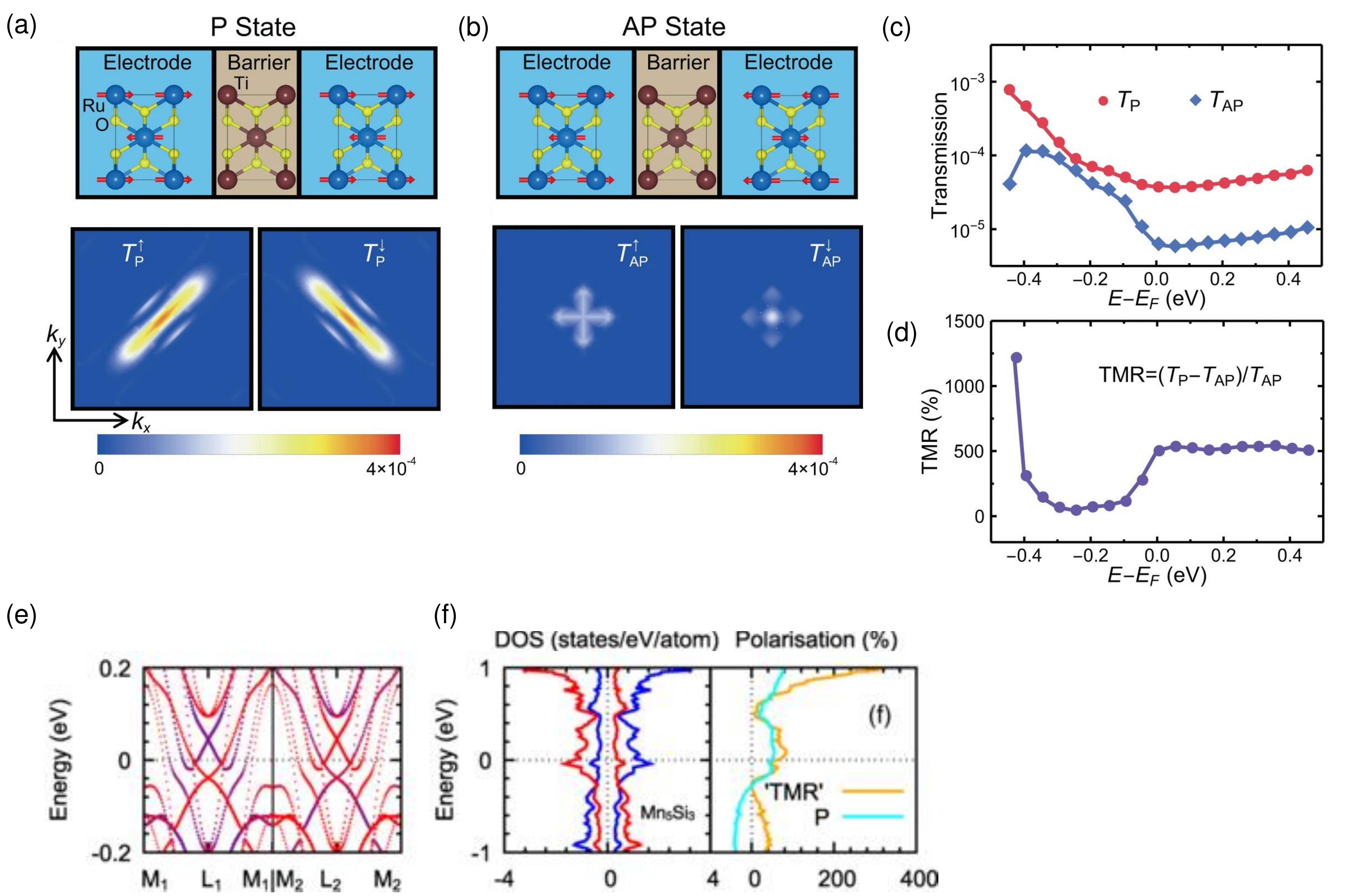}
	\caption{%
		(a)--(d) First-principles calculation of tunnel magnetoresistance (TMR) effect in the $\mathrm{RuO_{2}}(001)/\mathrm{TiO_{2}}(001)/\mathrm{RuO_{2}}$ magnetic tunnel junction.
		Figures are adapted from Ref.~\cite{Shao2021_NatCommun_12_7061} (\copyright Shao~\textit{et al.} (2021), licensed under CC BY 4.0).
		(a), (b) Magnetic states of the (a) parallel (P) and (b) antiparallel (AP) configurations and corresponding transmissions resolved by the momentum and spins.
		(c) Energy dependence of the total transmissions of parallel and antiparallel configurations.
		(d) Energy dependence of the TMR ratio.
		(e), (f) Results of first-principles calculation for a collinear antiferromagnet $\mathrm{Mn_{5}Si_{3}}$. 
		Figures are adapted from Ref.~\cite{Smejkal2022_PhysRevX_12_011028} (\copyright \v{S}mejkal~\textit{et al.} (2022), licensed under CC BY 4.0).
		(e) Spin and sublattice resolved band structure.
		(f) Spin and sublattice resolved density of states (left) and corresponding spin polarization and TMR effect.
	}
	\label{fig:shao_smejkal}
\end{figure*}
Next we discuss the TMR effect with antiferromagnets breaking the time-reversal symmetry~\cite{Chen2023_Nature_613_490,Shao2021_NatCommun_12_7061,Smejkal2022_PhysRevX_12_011028,Dong2022_PhysRevLett_128_197201,Qin2023_Nature_613_485,Cui2023_PhysRevB_108_024410,Gurung2023_arXiv_2306.03026,Jia2023_PhysRevB_108_104406,Jiang2023_PhysRevB_108_174439,Chi2024_PhysRevApplied_21_034038,Samanta2024_PhysRevB_109_174407,Shi2024_AdvMater_36_2312008,Zhu2024_ChinPhysLett_41_047502,Shao2024_npjSpintronics_2_13}.
Recently, it has been proposed that the antiferromagnets breaking the time-reversal symmetry can show various phenomena typically observed in ferromagnets~\cite{Noda2016_PhysChemChemPhys_18_13294,Okugawa2018_JPhysCondensMatter_30_075502,Naka2019_NatCommun_10_4305,Ahn2019_PhysRevB_99_184432,Smejkal2020_SciAdv_6_eaaz8809,Smejkal2022_PhysRevX_12_031042,Smejkal2022_PhysRevX_12_040501}.
These antiferromagnets is free from the problem on the interface discussed in the previous section.
\par
We will explain the antiferromagnetic TMR effect using the momentum dependent spin splitting.
Let us consider a collinear antiferromagnet breaking the time-reversal symmetry, 
which has been also called altermagnet recently~\cite{Smejkal2022_PhysRevX_12_031042,Smejkal2022_PhysRevX_12_040501}.
The band structures and the shapes of the Fermi surfaces are different between the majority and minority spin states;
the typical spin-resolved Fermi surfaces projected onto a two-dimensional plane are schematically shown in Fig.~\ref{fig:afm_tmr}(a).
We note that these spin splittings give the same amount of the total spin polarization when it is integrated over the whole Brillouin zone,
and the magnetization is compensated in total.
When the MTJs using such antiferromagnets are constructed, 
both electrodes have the same momentum-dependent spin splitting for the parallel configuration.
By contrast, for the antiparallel configuration,
two electrodes have the opposite momentum-dependent spin splitting each other (Fig.~\ref{fig:afm_tmr}(b)).
Therefore, the transmission for the parallel and antiparallel configurations have different momentum dependence,
which is still different even when integrated over the Brillouin zone.
Namely, a finite TMR effect can be realized in the antiferromagnetic MTJs.
\par
This type of the antiferromagnetic TMR effect owing to the spin splitting in the $\boldsymbol{k}$-space has been proposed by Shao~\textit{et al.}~\cite{Shao2021_NatCommun_12_7061}. 
They have presented that the MTJ with an altermagnetic candidate $\mathrm{RuO_{2}}$~\cite{Berlijn2017_PhysRevLett_118_077201,Ahn2019_PhysRevB_99_184432,Zhu2019_PhysRevLett_122_017202,Fedchenko2024_SciAdv_10_eadj4883,ruo2note} exhibits a finite TMR ratio by performing first-principles calculation.
$\mathrm{RuO_{2}}$ has two inequivalent Ru sites in a unit cell,
and due to the crystal structure,
its antiferromagnetic order breaks the time-reversal symmetry macroscopically.
They have performed first-principles calculation of the tunneling conductance for each of the parallel and antiparallel configurations in the $\mathrm{RuO_{2}}/\mathrm{TiO_{2}}/\mathrm{RuO_{2}}$ MTJ (Figs.~\ref{fig:shao_smejkal}(a) and~\ref{fig:shao_smejkal}(b)).
They have found the difference in the transmission between the parallel and antiparallel configurations (Fig.~\ref{fig:shao_smejkal}(c)) and corresponding finite TMR ratio (Fig.~\ref{fig:shao_smejkal}(d)) which will be comparable to the one observed in ferromagnetic MTJs.
\v{S}mejkal~\textit{et al.} have also focused on the altermagnet candidates such as $\mathrm{RuO_{2}}$ or $\mathrm{Mn_{5}Si_{3}}$~\cite{Smejkal2022_PhysRevX_12_011028}.
Based on the momentum dependent spin splitting properties shown in Fig.~\ref{fig:shao_smejkal}(e), 
they have shown that a finite TMR effect will be realized when these antiferromagnets are used as the electrodes of the MTJs (Fig.~\ref{fig:shao_smejkal}(f)).
\par
\begin{figure*}[tbh]
	\centering
	\includegraphics[width=160mm]{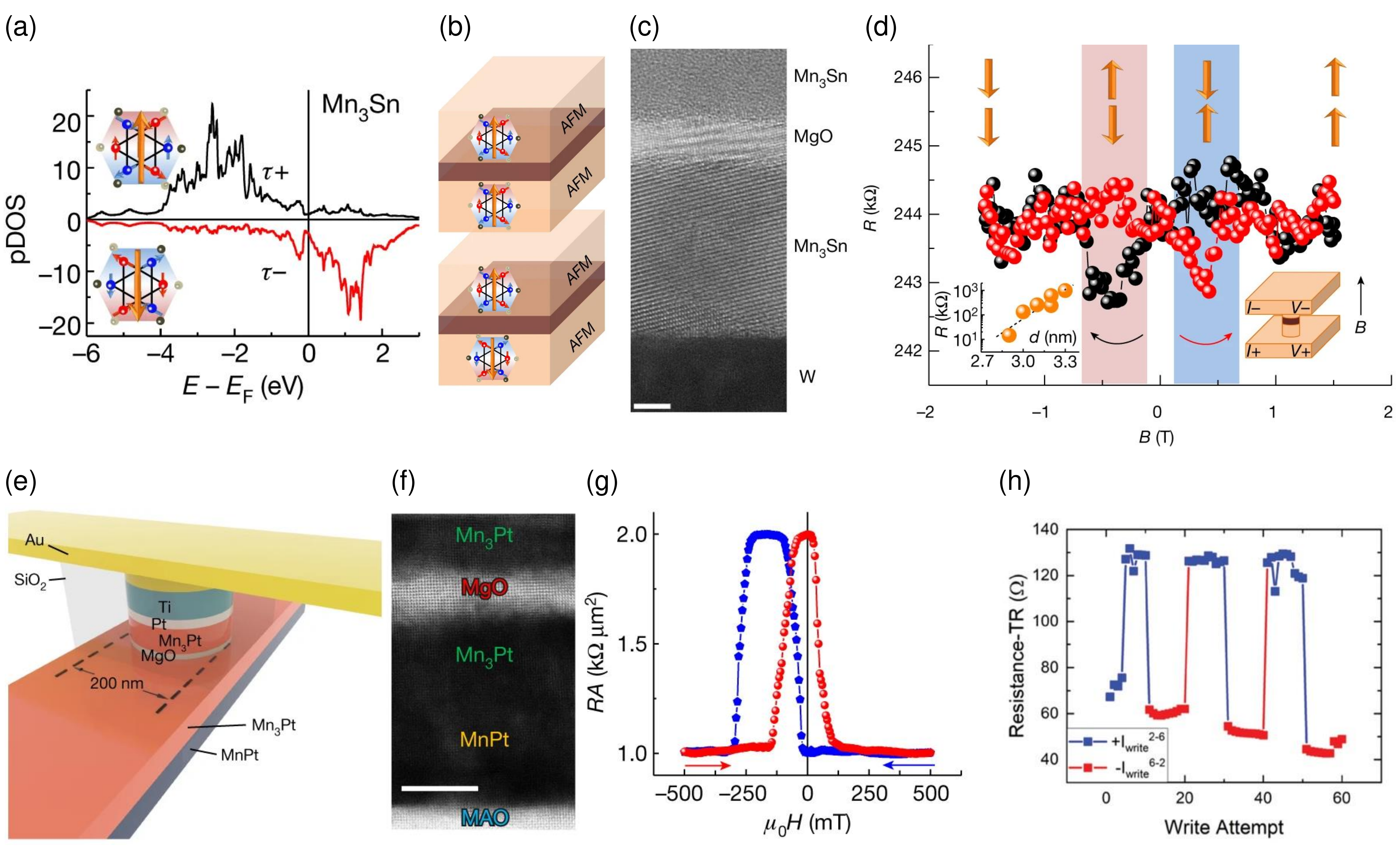}
	\caption{%
		(a)--(d) Tunnel magnetoresistance (TMR) effect in the $\mathrm{Mn_{3}Sn}/\mathrm{MgO}/\mathrm{Mn_{3}Sn}$ magnetic tunnel junction (MTJ).
		Figures are adapted from Ref.~\cite{Chen2023_Nature_613_490} (\copyright Chen~\textit{et al.} (2023), licensed under CC BY 4.0).
		(a) Projected density of states onto the two states with opposite cluster magnetic octupole moments.
		(b) Schematics of the parallel and antiparallel configurations of the $\mathrm{Mn_{3}Sn}/\mathrm{MgO}/\mathrm{Mn_{3}Sn}$ MTJ.
		(c) Transmission electron microscope (TEM) image of the $\mathrm{Mn_{3}Sn}/\mathrm{MgO}/\mathrm{Mn_{3}Sn}$ MTJ. Scale bar is 3~nm.
		(d) Magnetic field dependence of the tunneling resistance in the $\mathrm{Mn_{3}Sn}/\mathrm{MgO}/\mathrm{Mn_{3}Sn}$ MTJ.
		(e)--(g) TMR effect in the $\mathrm{Mn_{3}Pt}/\mathrm{MgO}/\mathrm{Mn_{3}Pt}$ MTJ. 
		Figures are adapted from Ref.~\cite{Qin2023_Nature_613_485} (\copyright Qin~\textit{et al.}, under exclusive licence to Springer Nature Limited 2023).
		(e) Schematics of the $\mathrm{Mn_{3}Pt}/\mathrm{MgO}/\mathrm{Mn_{3}Pt}$ MTJ.
		(f) TEM image of the $\mathrm{Mn_{3}Pt}/\mathrm{MgO}/\mathrm{Mn_{3}Pt}$ MTJ. Scale bar is 5~nm.
		(g) Magnetic field dependence of the tunneling resistance in the $\mathrm{Mn_{3}Pt}/\mathrm{MgO}/\mathrm{Mn_{3}Pt}$ MTJ.
		(h) Changes of the resistance by applying an electric current in the $\mathrm{Mn_{3}Pt}/\mathrm{Al_{2}O_{3}}/\mathrm{Mn_{3}Pt}$ MTJ.
		Figure is adopted from Ref.~\cite{Shi2024_AdvMater_36_2312008} (\copyright Shi~\textit{et al.} (2024)).
	}
	\label{fig:chen_qin_2023}
\end{figure*}
The TMR effect with the antiferromagnets breaking time-reversal symmetry has been discussed using noncollinear antiferromagnets as well as the altermagnets or the collinear antiferromagnets,
and been observed in experiments.
One of such noncollinear antiferromagnets is $\mathrm{Mn_{3}Sn}$ with a hexagonal crystal structure.
The $\mathrm{Mn}$ atoms in $\mathrm{Mn_{3}Sn}$ form the kagome lattice and the magnetic moments take the inverse triangular structure which breaks the time-reversal symmetry~\cite{Tomiyoshi1982_JPhysSocJpn_51_803,Nagamiya1982_SolidStateCommun_42_385,Brown1990_JPhysCondensMatter_2_9409}.
This antiferromagnetic state can be characterized by the cluster magnetic octupole moment~\cite{Suzuki2017_PhysRevB_95_094406,Suzuki2019_PhysRevB_99_174407} as described by the projected density of states onto the cluster magnetic octupole states shown in Fig.~\ref{fig:chen_qin_2023}(a).
This magnetic structure has led to the observation of various ferromagnetic-like transport phenomena~\cite{Nakatsuji2015_Nature_527_212,Ikhlas2017_NatPhys_13_1085,Higo2018_NatPhoton_12_73,Higo2022_JMagnMagnMater_564_170176,Nakatsuji2022_AnnuRevCondensMatterPhys_13_119}.
Chen~\textit{et al.} found a finite TMR effect in an all-antiferromagnetic tunnel junction based on $\mathrm{Mn_{3}Sn}$ (Figs.~\ref{fig:chen_qin_2023}(b) and \ref{fig:chen_qin_2023}(c))~\cite{Chen2023_Nature_613_490}.
They experimentally observed the difference in the tunnel resistance between when the cluster octupole moments of two magnetic layers align parallelly and antiparallelly (Fig.~\ref{fig:chen_qin_2023}(d)), 
as well as first-principles calculation showing a finite TMR effect with $\mathrm{Mn_{3}Sn}$.
Theoretical approaches to the $\mathrm{Mn_{3}Sn}$-based MTJ are also performed by Dong~\textit{et al.}~\cite{Dong2022_PhysRevLett_128_197201},
where they show the relative angle dependence of the cluster magnetic octupole moments between two electrodes in addition to the parallel and antiparallel alignments.
\par
Another noncollinear antiferromagnet showing the TMR effect is $\mathrm{Mn_{3}Pt}$ with a cubic crystal structure. 
The $\mathrm{Mn}$ ions in the $(111)$-plane of the $\mathrm{Mn_{3}Pt}$ form the kagome lattice and the antiferromagnetic structure breaks the time-reversal symmetry macroscopically~\cite{Kren1966_PhysLett_20_331,Kren1967_JApplPhys_38_1265,Kren1968_PhysRev_171_574}.
Qin~\textit{et al.} observed a finite TMR effect in the $\mathrm{Mn_{3}Pt}/\mathrm{MgO}/\mathrm{Mn_{3}Pt}$ MTJ (Fig.~\ref{fig:chen_qin_2023}(e) and Fig.~\ref{fig:chen_qin_2023}(f))~\cite{Qin2023_Nature_613_485}.
They utilize the exchange bias effect between $\mathrm{Mn_{3}Pt}$ and $\mathrm{MnPt}$ and present that the TMR effect will reach even around 100\%~(Fig.~\ref{fig:chen_qin_2023}(g)).
A finite TMR effect in the $\mathrm{Mn_{3}Pt}$-based MTJ has been also observed by Shi~\textit{et al.} in the $\mathrm{Mn_{3}Pt}/\mathrm{Al_{2}O_{3}}/\mathrm{Mn_{3}Pt}$ MTJ~\cite{Shi2024_AdvMater_36_2312008}.
They have performed the switching of the magnetic moments of $\mathrm{Mn_{3}Pt}$ by applying the electric current instead of the magnetic field (Fig.~\ref{fig:chen_qin_2023}(h)).
This result will lead to the switching of the antiferromagnetic MTJs by the electric current as well as the magnetic field,
which will be more suitable for further application in antiferromagnetic spintronics devices.
\par
The momentum dependent spin splitting in the time-reversal symmetry breaking antiferromagnet also expects us to realize the antiferromagnet/insulator/ferromagnet MTJ,
in addition to the all antiferromagnetic tunnel junctions discussed above.
The TMR effect in the $\mathrm{Fe}/\mathrm{MgO}/\mathrm{Mn_{3}Sn}$ tunnel junction has been experimentally observed by Chen~\textit{et al.}~\cite{Chen2023_Nature_613_490}.
Chou \textit{et al.} have shown the TMR effect in the $\mathrm{CoFeB}/\mathrm{MgO}/\mathrm{Mn_{3}Sn}$ MTJ at low temperature~\cite{Chou2024_NatCommun_15_7840}.
Theoretical calculations have been performed in $\mathrm{RuO_{2}}/\mathrm{TiO_{2}}/\mathrm{CrO_{2}}$ MTJ~\cite{Chi2024_PhysRevApplied_21_034038,Samanta2024_PhysRevB_109_174407},
where they utilize the half-metallicity of the ferromagnet $\mathrm{CrO_{2}}$~\cite{Schwarz1986_JPhysFMetPhys_16_L211,Korotin1998_PhysRevLett_80_4305}.
\section{A real-space approach to antiferromagnetic tunnel magnetoresistance effect: local density of states}
\label{sec:tmr_ldos}
In the previous section, we discussed the antiferromagnetic TMR effect from the viewpoint of the spin splitting in the momentum space.
In this section, based on Ref.~\cite{Tanaka2023_PhysRevB_107_214442}, 
we discuss a real-space approach to capture the TMR effect, particularly using the LDOS.
\par
We show that the LDOS inside the barrier will work as an indicator of the TMR effect.
In the original Julliere model, 
the product of the bulk density of states is used to estimate the ferromagnetic TMR effect as discussed in Sec.~\ref{subsec:afm_tmr_ferro}.
In a similar manner, here we present that the product of the LDOS of the barrier region can trace the TMR effect qualitatively.
In contrast to the Julliere model which does not describe the antiferromagnetic TMR effect,
this method with the LDOS will work in the ferrimagnetic and antiferromagnetic MTJs as well as the ferromagnetic MTJs.
\subsection{Local density of states}
Similarly to the Julliere model (Eq.~(\ref{eq:dos_product})),
we can consider the product of the LDOS inside the barrier given as
\begin{eqnarray}
\tau_{\text{LDOS}} = \sum_{i = 1}^{n_{\text{a}}} \sum_{\sigma} d_{\text{L}, i, \sigma} d_{\text{R}, i, \sigma}.
\label{eq:ldos_product}
\end{eqnarray}
Here, $\text{L}$ and $\text{R}$ is the indices of the layers we focus on,
and $d_{\text{L/R}, i, \sigma}$ is the LDOS at the $i$-th atom on the $\text{L/R}$ layer.
The $i$-th atom in the L and R layers share the same in-plane coordinates perpendicular to the conducting path.
The number of atoms in the layer is $n_{\text{a}}$.
\subsection{Lattice model calculation}
\label{subsec:tmr_ldos_model}
By a lattice model calculation, we show that $\tau_{\text{LDOS}}$ qualitatively captures the transmission property.
To mimic the MTJ, we consider the two-dimensional square lattice consisting of three parts;
the left and right electrodes semi-infinitely extend in the conducting direction,
and the barrier region between two electrodes has a length of $L$ in the conducting direction.
Each of the two electrodes and the barrier has a width of $W$ in the direction perpendicular to the conducting path (Fig.~\ref{fig:prb_ferro}(a)).
We deal with the following Hamiltonian on this lattice given as;
\begin{eqnarray}
	\mathcal{H} 
= &\sum_{i} \varepsilon_{i} n_{i} 
	- t \sum_{\braket{i, j}} \left( c_{i, \sigma}^{\dag} c_{j, \sigma} + \text{h.c.} \right) \nonumber \\
& - J \sum_{i \in \text{electrode}, \alpha, \beta} \left( \boldsymbol{s}_{i} \cdot \boldsymbol{\sigma} \right)^{\alpha\beta} c_{i, \alpha}^{\dag} c_{i, \beta}.
\end{eqnarray}
Here, $c_{i, \alpha}^{\dag}/c_{i, \alpha}$ ($\alpha = \uparrow, \downarrow$) is the creation/annihilation operator of an electron with spin-$\alpha$ on site-$i$,
$n_{i} = \sum_{\alpha} c_{i, \alpha}^{\dag} c_{i, \alpha}$,
and $\boldsymbol{\sigma} = {}^{t}( \sigma^{x} \ \sigma^{y} \ \sigma^{z} )$ is the $2\times 2$ Pauli matrices.
The on-site energy is $\varepsilon_{i}$.
The electron hopping between site-$i$ and $j$ is $t$;
the nearest-neighbor hopping is considered here, which is expressed by $\braket{i, j}$.
The localized spin moment on the $i$-th site in the electrode is $\boldsymbol{s}_{i}$,
and $J$ is the magnetic coupling constant.
We set $\varepsilon_{i} = 10$ for the barrier region and $t = 1$.
We take the system size of the barrier region as $L = 8$ and $W = 160$,
and impose the open boundary condition on the perpendicular direction to the conducting path.
\par
We use the \textsc{kwant} package for the simulation,
where the quantum ballistic transport is calculated~\cite{Groth2014_NewJPhys_16_063065}.
\subsection{Results}
\subsubsection{Ferromagnetic electrode}
\begin{figure*}[tbh]
	\centering
	\includegraphics[width=120mm]{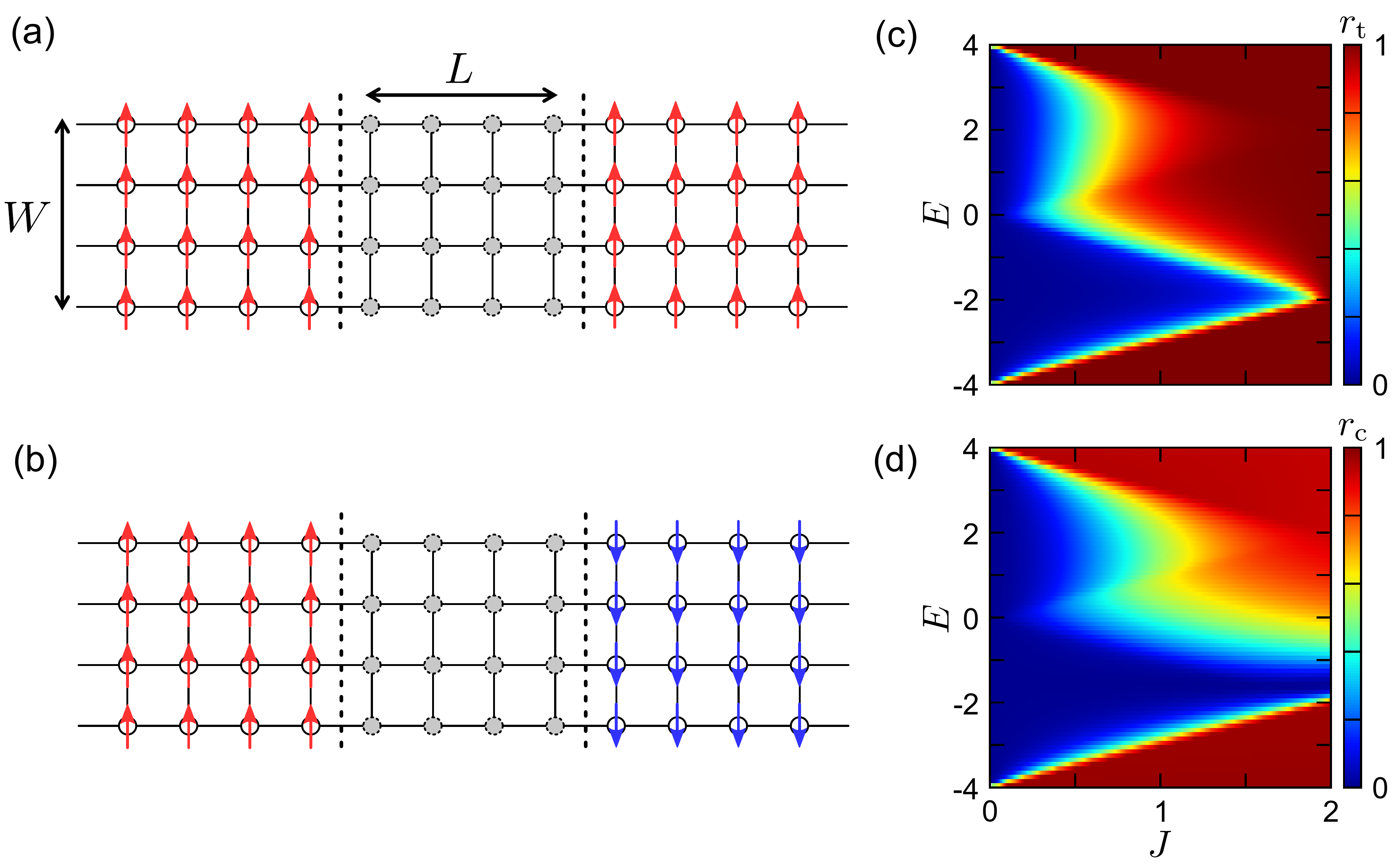}
	\caption{%
		(a), (b) Schematic figures of the magnetic tunnel junction with ferromagnetic electrodes for the (a) parallel and (b) antiparallel configurations.
		(c) Tunnel magnetoresistance ratio, $r_{\text{t}}$, defined by Eq.~(\ref{eq:tmr_ratio_model}), with respect to $J$ and $E$.
		(d) The ratio defined by the product of the local density of states, $r_{\text{c}}$ (Eq.~(\ref{eq:ldos_ratio_model})).
		Figures are adopted from Ref.~\cite{Tanaka2023_PhysRevB_107_214442} (\copyright American Physical Society (2023)).
	}
	\label{fig:prb_ferro}
\end{figure*}
We first deal with the ferromagnetic tunnel junction to test the applicability of the method with LDOS.
The left electrode has $\boldsymbol{s}_{i} = (0, 0, 1)$ for all sites.
In the right electrode, 
every site has $\boldsymbol{s}_{i} = (0, 0, 1)$ for the parallel configuration,
and $\boldsymbol{s} = (0, 0, -1)$ for the antiparallel configuration.
The parallel and antiparallel configurations of the ferromagnetic MTJ are schematically shown in Fig.~\ref{fig:prb_ferro}(a);
\par
We present the TMR ratio, $r_{\text{t}}$, on the plane of $J$ and $E$ in Fig.~\ref{fig:prb_ferro}(b).
Here the TMR ratio is given as
\begin{eqnarray}
	r_{\text{t}}
= \frac{T_{\text{P}}-T_{\text{AP}}}{T_{\text{P}}+T_{\text{AP}}},
\label{eq:tmr_ratio_model}
\end{eqnarray}
where $T_{\text{P}}$ and $T_{\text{AP}}$ are the transmission for the parallel and antiparallel alignments of the MTJ.
When the magnetic interaction $J = 0$, the whole system is nonmagnetic,
and the TMR ratio becomes zero.
By introducing a finite magnetic interaction $J$,
the parallel and antiparallel configurations are distinguished, 
and the TMR ratio becomes finite.
\par
In Fig.~\ref{fig:prb_ferro}(c), we show the ratio, $r_{\text{c}}$, defined by the LDOS, as
\begin{eqnarray}
	r_{\text{c}}
= \frac{\tau_{\text{LDOS, P}}-\tau_{\text{LDOS, AP}}}{\tau_{\text{LDOS, P}}+\tau_{\text{LDOS, AP}}},
\label{eq:ldos_ratio_model}
\end{eqnarray}
where $\tau_{\text{LDOS, P/AP}}$ is $\tau_{\text{LDOS}}$ for the parallel/antiparallel configuration.
In this model calculations, 
we calculate $\tau_{\text{LDOS}}$ using the $L/2$-th (fourth) layer and $L/2 + 1$-th (fifth) layer~\cite{latticemodelnote}.
By comparing Fig.~\ref{fig:prb_ferro}(b) and Fig.~\ref{fig:prb_ferro}(c),
we find that $r_{\text{t}}$ and $r_{\text{c}}$ are qualitatively similar to each other.
Namely, we can estimate the TMR property using the LDOS inside the barrier.
\subsubsection{Ferrimagnetic electrode}
\begin{figure*}[tbh]
	\centering
	\includegraphics[width=160mm]{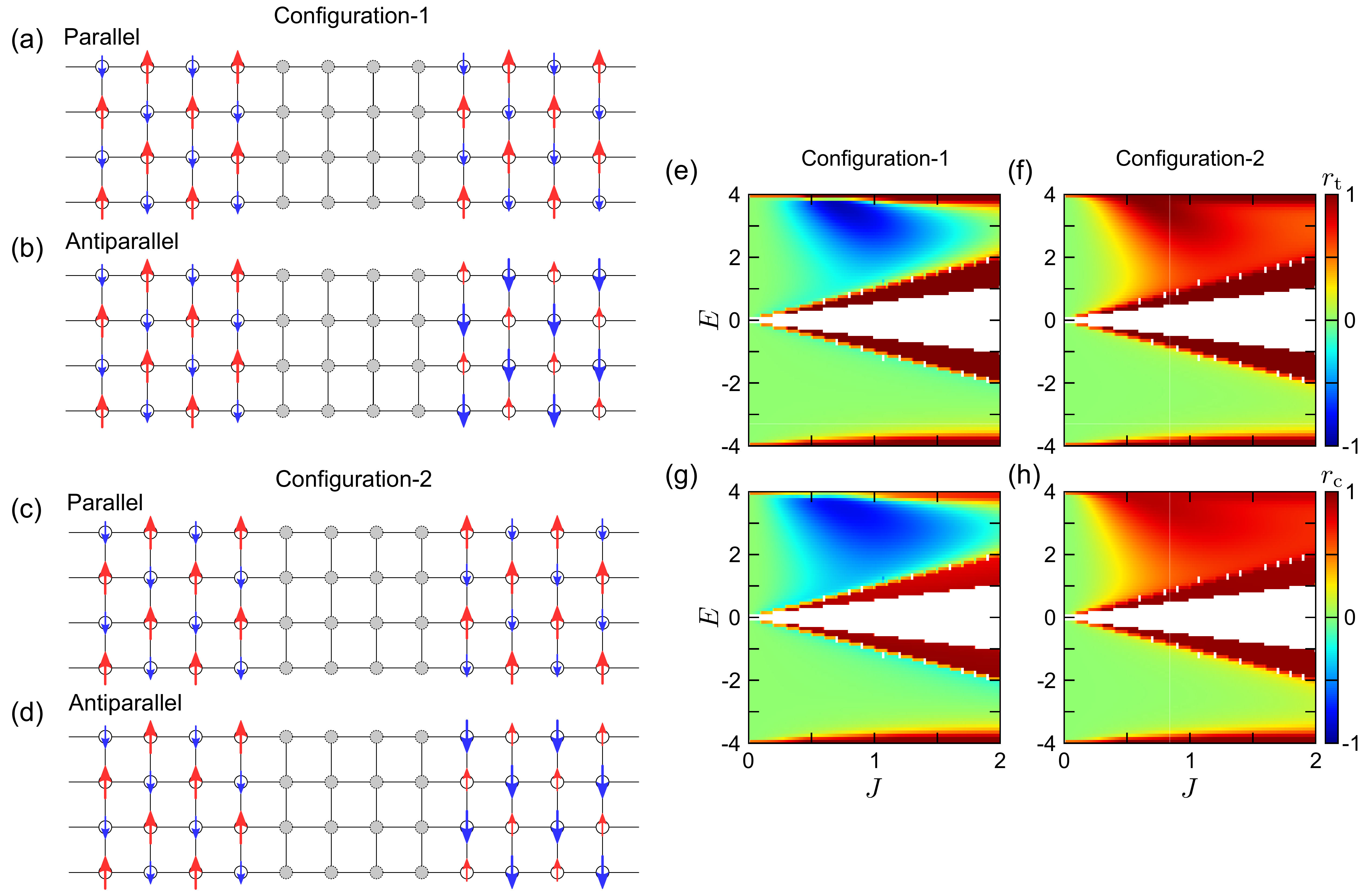}
	\caption{%
		Simulation results for the ferrimagnetic tunnel junctions with $\boldsymbol{s}_{\text{A}} = (0, 0, 1)$ and $\boldsymbol{s}_{\text{B}} = (0, 0, -0.5)$.
		(a), (b) Colormaps of the tunnel magnetoresistance ratio, $r_{\text{t}}$ (Eq.~(\ref{eq:tmr_ratio_model})), as a function of $J$ and $E$.
		(e), (f) Ratio defined by the product of the local density of states  (LDOS) inside the barrier, $r_{\text{c}}$ (Eq.~(\ref{eq:ldos_ratio_model})).
		Figures are adopted from Ref.~\cite{Tanaka2023_PhysRevB_107_214442} (\copyright American Physical Society (2023)).
	}
	\label{fig:prb_ferri}
\end{figure*}
Next we present the results of the simulations for the MTJs with ferrimagnetic electrodes.
Unlike the ferromagnetic MTJs, 
we have multiple sets of the parallel and antiparallel configurations for the ferrimagnetic MTJs,
which is owing to the sublattice degrees of freedom of ferrimagnets.
Here we consider the ferrimagnet with two sublattices, A and B, 
with $\boldsymbol{s}_{\text{A}} = (0, 0, 1)$ and $\boldsymbol{s}_{\text{B}} = (0, 0, -0.5)$.
In one configuration, we call configuration-1, 
the sites in the different sublattices sandwich the barrier as shown in Figs.~\ref{fig:prb_ferri}(a) and \ref{fig:prb_ferri}(b).
In the other configuration, we call configuration-2, 
the sites in the same sublattices are facing to each other across the barrier region (see Figs.~\ref{fig:prb_ferri}(c) and \ref{fig:prb_ferri}(d)).
The parallel and antiparallel configurations of the ferrimagnetic MTJ are given by the relative directions of the spin moments in the same sublattice;
when the spins in the A-sublattice of the left and right electrodes align parallelly (antiparallelly), 
the MTJ takes the parallel (antiparallel) configuration.
\par
We show the TMR ratio $r_{\text{t}}$ (Eq.~(\ref{eq:tmr_ratio_model})) with respect to $J$ and $E$ for the configuration-1 and 2 in Figs.~\ref{fig:prb_ferri}(e) and \ref{fig:prb_ferri}(f), respectively.
The ratio calculated by the LDOS, $r_{\text{c}}$ (Eq.~(\ref{eq:ldos_ratio_model})) for the configurtation-1 and 2 are respectively shown in Figs.~\ref{fig:prb_ferri}(g) and \ref{fig:prb_ferri}(h).
By comparing $r_{\text{t}}$ and $r_{\text{c}}$ for each of the two configurations,
we confirm that $\tau_{\text{LDOS}}$ can trace the qualitative TMR property also for the ferrimagnetic MTJs.
\par
We remark that the interfacial spins roughly determine the TMR property.
In the configuration-1, spins at the interface of the left and right electrodes align antiparallelly (parallelly) in the parallel (antiparallel) configuration (Figs.~\ref{fig:prb_ferri}(a) and \ref{fig:prb_ferri}(b)).
This leads to an inverse TMR effect, namely, $r_{\text{t}} < 0$, in some parameter regions of the configuration-1.
On the other hand, in the configuration-2, 
interfacial spins of the two electrodes across the barrier region take parallel (antiparallel) in the parallel (antiparallel) directions (Figs.~\ref{fig:prb_ferri}(c) and \ref{fig:prb_ferri}(d)),
and thus $r_{\text{t}} > 0$ basically holds.
\subsubsection{Antiferromagnetic electrode}
\begin{figure}[tbh]
	\centering
	\includegraphics[width=80mm]{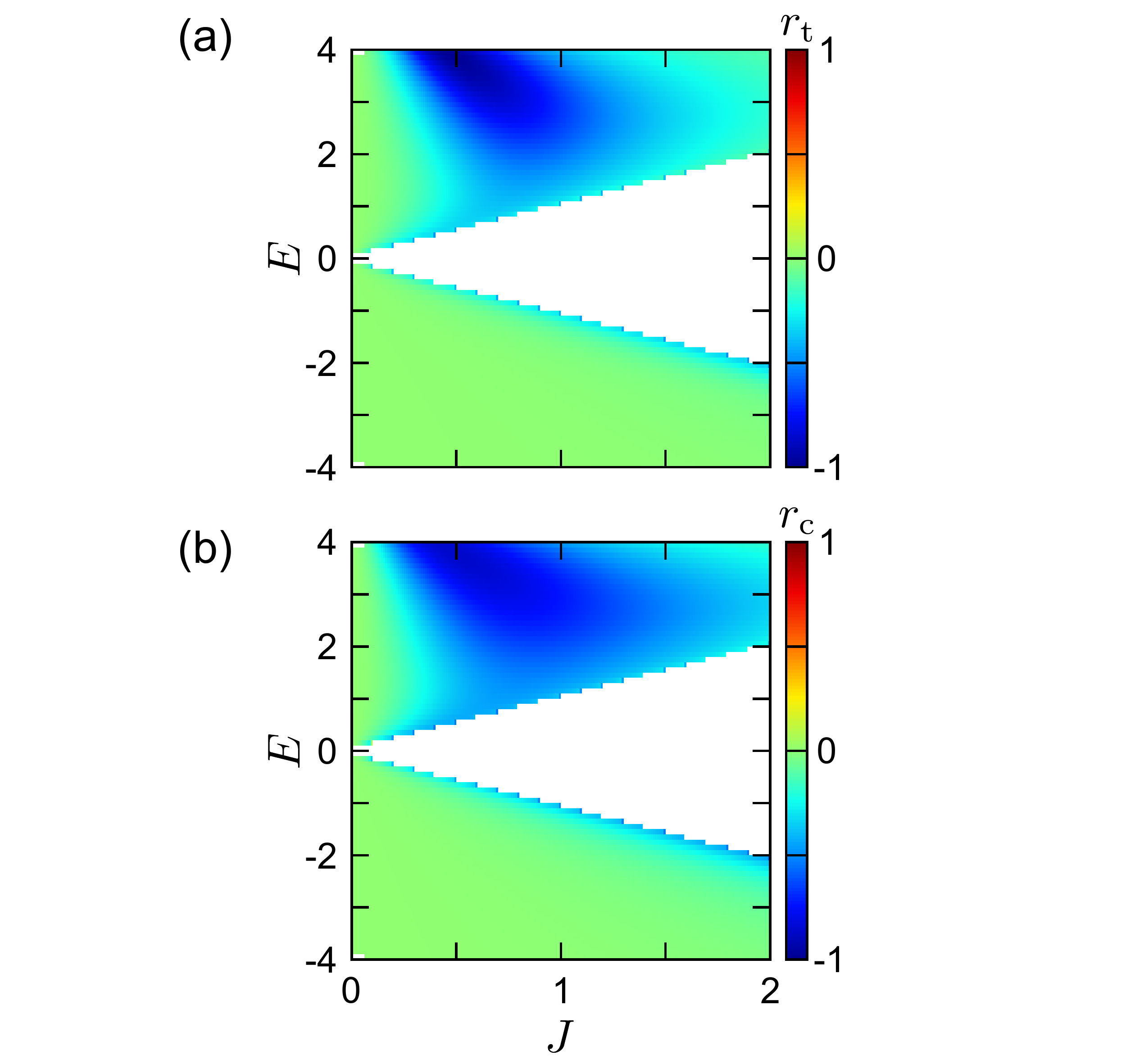}
	\caption{%
		Simulation results for the antiferromagnetic tunnel junction with $\boldsymbol{s}_{\text{A}} = (0, 0, 1)$ and $\boldsymbol{s}_{\text{B}} = (0, 0, -1)$.
		(a) Colormap of the tunnel magnetoresistance ratio $r_{\text{t}}$ as a function of $J$ and $E$.
		(b) Ratio defined by the product of the local density of states (LDOS) inside the barrier, $r_{\text{c}}$.
		Figures are adopted from Ref.~\cite{Tanaka2023_PhysRevB_107_214442} (\copyright American Physical Society (2023)).
	}
	\label{fig:prb_antiferro}
\end{figure}
Finally, we discuss the TMR effect with antiferromagnetic electrodes.
We deal with the antiferromagnetic electrodes by taking $\boldsymbol{s}_{\text{A}} = (0, 0, 1)$ and $\boldsymbol{s}_{\text{B}} = (0, 0, -1)$ in the configuration-1 of the above ferrimagnetic electrode.
We show the TMR ratio $r_{\text{t}}$ (Eq.~(\ref{eq:tmr_ratio_model})) and the ratio defined by the LDOS $r_{\text{c}}$ (Eq.~(\ref{eq:ldos_ratio_model})) in Figs.~\ref{fig:prb_antiferro}(a) and \ref{fig:prb_antiferro}(b), respectively.
Also, in the case with the antiferromagnetic electrode, 
we confirm that the TMR ratio is qualitatively captured by the product of the LDOS inside the barrier.
\par
We should note that we have considered the case where the time-reversal symmetry is preserved here;
configuration-1 and configuration-2 are equivalent with each other.
Hence the TMR effect will be cancelled out if we consider the other configuration similarly to the ferrimagnetic MTJs.
However, this approach itself will be applicable to the TRS-broken antiferromagnets.
Namely, the configuration-1 and 2 are distinguished when we use the antiferromagnets breaking the TRS.
In that case, a finite TMR will be realized.
\subsection{Related approaches to the tunnel magnetoresistance effect with local density of states}
Here we have introduced a real-space LDOS approach to the TMR effect.
We have utilized the LDOS inside the barrier which has the information on the tunneling decay and the details of the electrodes and the barrier.
We remark that another real-space approach using the LDOS was proposed for the TMR effect by Tsymbal~\textit{et al.}~\cite{Tsymbal2005_JApplPhys_97_10C910,Tsymbal2007_ProgMaterSci_52_401};
they evaluated the transmission with the interfacial LDOS and the exponential decay factor.
In addition, the momentum-resolved LDOS has been utilized to analyze the TMR effect~\cite{Belashchenko2004_PhysRevB_69_174408,Stamenova2017_PhysRevB_95_060403,BaezFlores2024_arXiv_2404.08103}.
Our approach and these approaches introduced here will be complementary to each other.
\section{Application of the approach with local density of states to first-principles calculation: $\mathrm{Fe/MgO/Fe}$ tunnel junction}
\label{sec:fe_mgo_fe}
In this section, we discuss an application of the method discussed above to first-principles calculation.
We have discussed the role of the LDOS for evaluating the TMR effect using the lattice model.
However, to utilize this method for designing the MTJ, particularly the antiferromagnetic MTJs,
we should demonstrate that it works also for the materials.
Here, we discuss the Fe(001)/MgO(001)/Fe MTJ as an example, 
which is one of the representative MTJ to show a distinctive TMR effect~\cite{Butler2001_PhysRevB_63_054416,Mathon2001_PhysRevB_63_220403,Yuasa2004_JpnJApplPhys_43_L588,Parkin2004_NatMater_3_862,Yuasa2004_NatMater_3_868,Tusche2005_PhysRevLett_95_176101,Waldron2006_PhysRevLett_97_226802,Matsumoto2007_ApplPhysLett_90_252506,Heiliger2007_PhysRevB_77_224407,Heiliger2007_JMagnMagnMater_316_478,Rungger2007_JMagnMagnMater_316_481,Wang2008_PhysRevB_78_180411,Bose2010_PhysRevB_82_014412,Raza2011_JApplPhys_109_023705,Autes2011_PhysRevB_84_134404,Masuda2021_PhysRevB_104_L180403,Scheike2021_ApplPhysLett_118_042411}.
\subsection{System and Method}
\begin{figure*}[tbh]
	\centering
	\includegraphics[width=160mm]{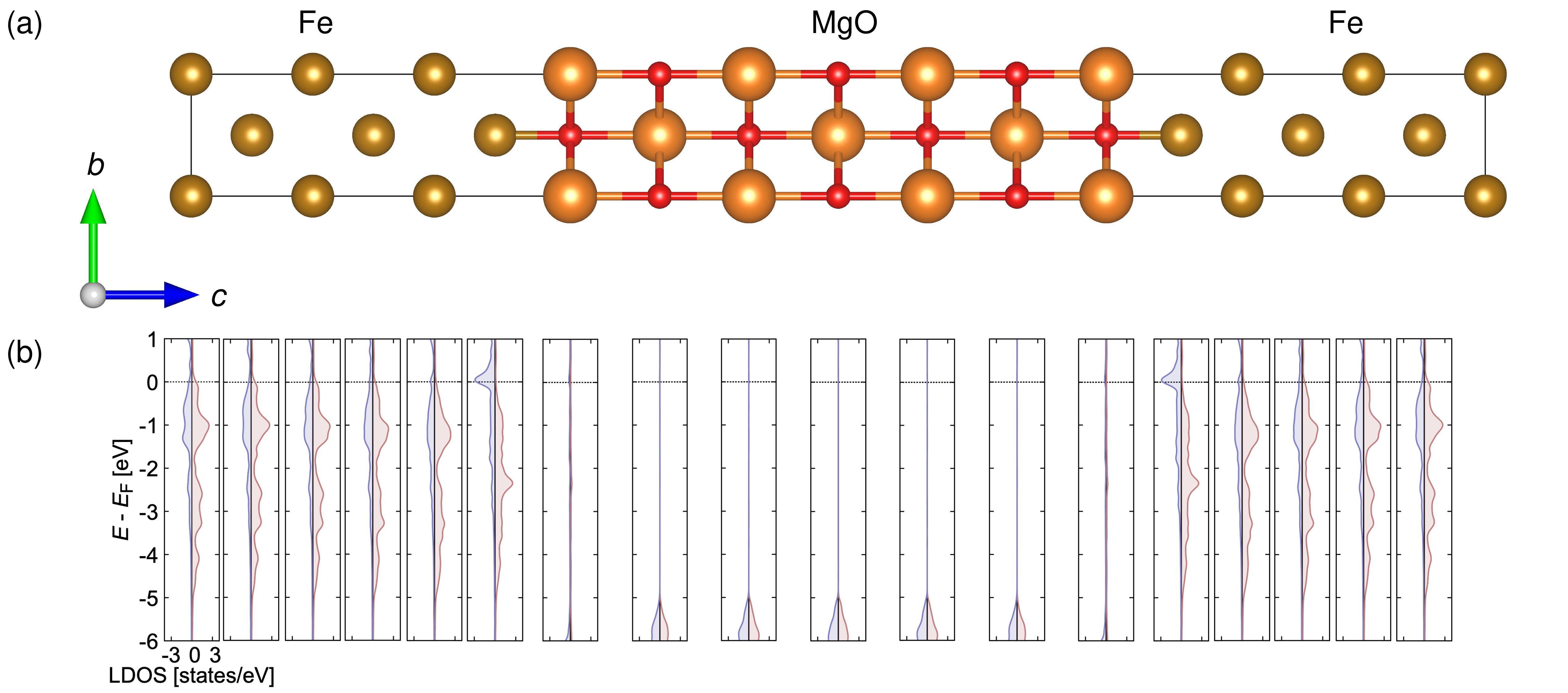}
	\caption{%
		(a) Crystal structure of the scattering region for the Fe/MgO/Fe tunnel junction with seven monolayers of MgO barrier.
		(b) Local density of states at each layer of the scattering region for the parallel configuration.
	}
	\label{fig:femgofe_mtj}
\end{figure*}
We use bcc-Fe as the magnetic electrode, whose lattice constant we use is $2.8665$~\AA.
The insulating barrier consists of MgO, 
whose in-plane lattice constant is taken the same with that of bcc-Fe,
and the $c$-axis lattice constant of MgO is set as $4.2113$~\AA.
The interface between Fe and MgO is given by the average of the lattice constants of Fe and MgO.
\par
We divide the whole MTJ into three regions, namely, the left and right leads and the scattering region in between,
and perform the electronic structure calculation for each of the three parts.
The left and right leads are Fe.
The scattering region has seven monolayers (MLs) of MgO sandwiched by six and five MLs of Fe as shown in Fig.~\ref{fig:femgofe_mtj}(a).
For the calculation of the MTJ with antiparallel configuration, 
we deal with the doubled scattering region along the conducting path for the smooth connection;
we attach the supercell which consists of the scattering region with the magnetic moments inverted to the original scattering region.
The doubled scattering region is cut in half, and return to the original scattering region when calculating the transmission.
\par
To perform the density functional theory calculation~\cite{Hohenberg1964_PhysRev_136_B864,Kohn1965_PhysRev_140_A1133},
we use the \textsc{Quantum ESPRESSO} (\textsc{QE}) package~\cite{Giannozzi2009_JPhysCondensMatter_21_395502,Giannozzi2017_JPhysCondensMatter_29_465901}.
We use the ultrasoft-type pseudopotential obtained from \textsc{pslibrary}~\cite{DalCorso2014_ComptMaterSci_95_337}.
The exchange correlation is taken in by the Perdew--Burke--Ernzerhof (PBE) generalized gradient approximation (GGA)~\cite{Perdew1996_PhysRevLett_77_3865}.
The cutoff of the energy for the wave-function is 100~Ry, and that for the charge-density is 600~Ry.
We perform self-consistent field (scf) calculation with the $\boldsymbol{k}$-point mesh of $20 \times 20 \times 20$ for the lead and $20 \times 20 \times 1$ for the scattering region.
The effect of the spin-orbit coupling (SOC) is neglected.
\par
After the scf calculations of the leads and the scattering region, 
we attach the three parts to construct the MTJ, and calculate the transmission.
We use the \textsc{pwcond} routine contained in the \textsc{QE} package to calculate the transmission~\cite{Smogunov2004_PhysRevB_70_045417,DalCorso2005_PhysRevB_71_115106,DalCorso2006_PhysRevB_74_045429},
which takes the scattering approach~\cite{Choi1999_PhysRevB_59_2267}.
The tunnel conductance is calculated by the Landauer--B\"{u}ttiker formula~\cite{Landauer1957_IBMJResDev_1_3,Landauer1970_PhilMag_21_863,Buttiker1986_PhysRevLett_57_1761,Buttiker1988_IBMJResDevelop_32_317}, which relates the transmission to the conductance as $G = (e^{2}/h)T$.
The transmission is calculated as
\begin{eqnarray}
T = \frac{1}{N_{\boldsymbol{k}_{\parallel}}} \sum_{\boldsymbol{k}_{\parallel}, \sigma} T_{\sigma}(\boldsymbol{k}_{\parallel}),
\end{eqnarray}
where $T_{\sigma}(\boldsymbol{k}_{\parallel})$ is the partial transmission of the electron with the spin-$\sigma$ and the in-plane momentum of $\boldsymbol{k}_{\parallel}$.
The in-plane $\boldsymbol{k}$-mesh in the transmission calculation, $N_{\boldsymbol{k}_{\parallel}}$, is $401 \times 401$.
\par
For the calculation of LDOS, we perform nscf calculation of the scattering region after the scf calculation. 
The $\boldsymbol{k}$-point mesh of nscf calculation is $40 \times 40 \times 1$.
\subsection{Results and discussions}
\begin{figure}[tbh]
	\centering
	\includegraphics[width=80mm]{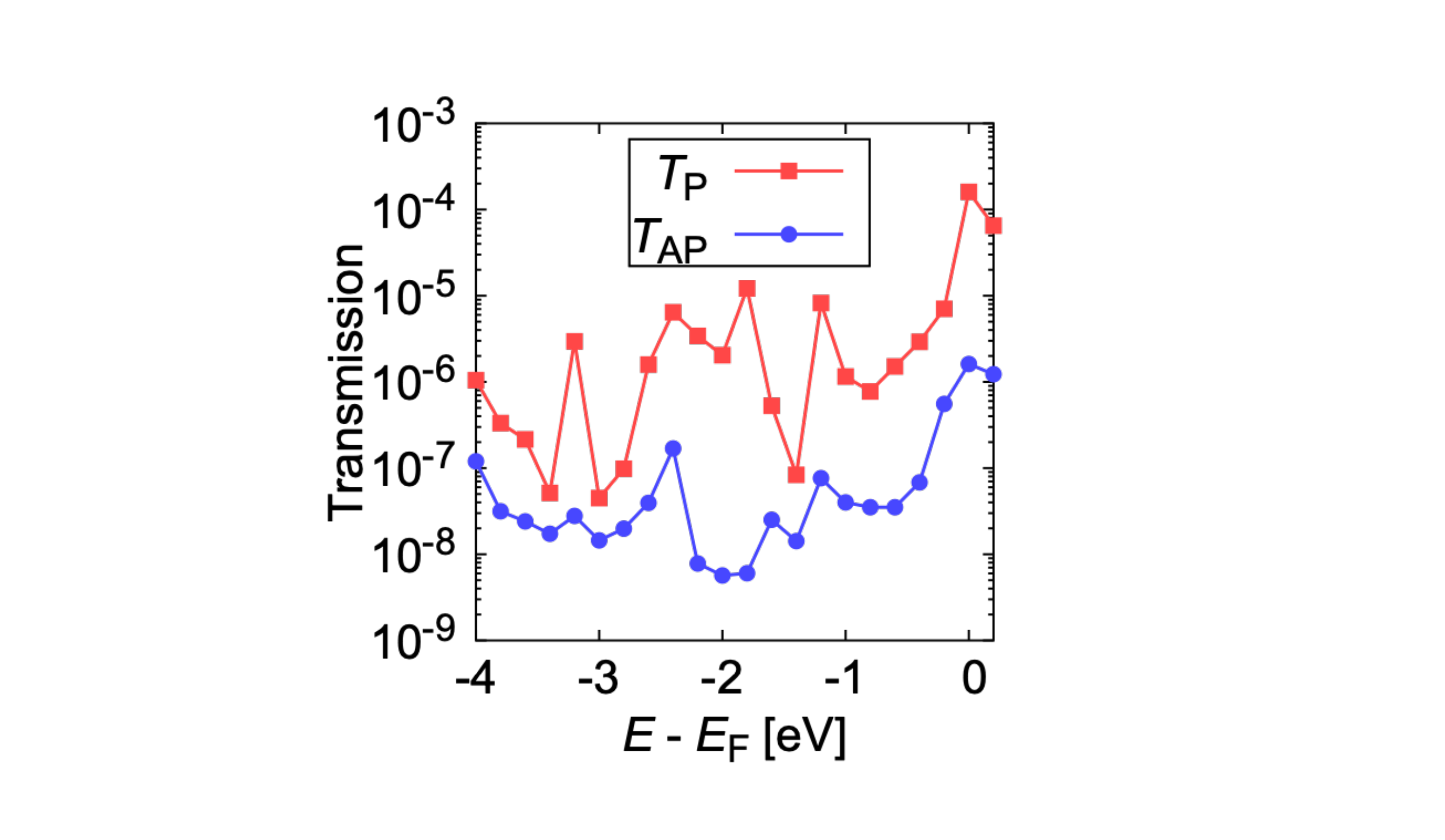}
	\caption{%
		Total transmissions for the parallel and antiparallel configurations of the Fe/MgO/Fe tunnel junction with respect to the energy.
	}
	\label{fig:femgofe_transmission}
\end{figure}
The spin resolved LDOS of each layer of the scattering region with the parallel configuration is shown in Fig.~\ref{fig:femgofe_mtj}(b).
We confirm that the LDOS inside MgO is small enough in the energy region we focus on,
which shows that the transport discussed here is a tunneling one.
\par
The total transmission of the parallel and antiparallel configurations with respect to the chemical potential are shown in Fig.~\ref{fig:femgofe_transmission}(a).
In the whole energy region, we find that $T_{\text{P}}$ takes a larger value than $T_{\text{AP}}$.
\begin{figure}[tbh]
	\centering
	\includegraphics[width=80mm]{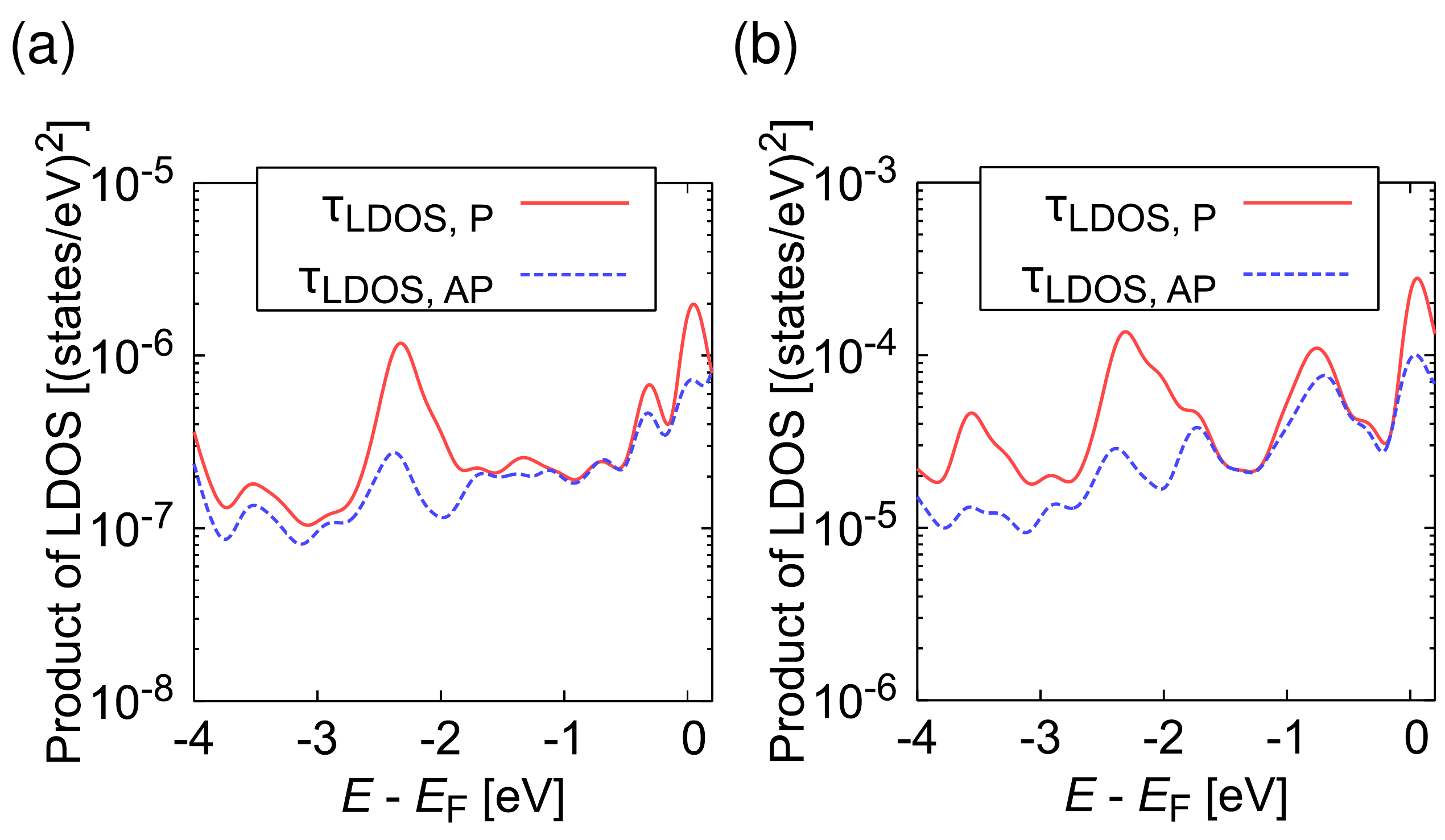}
	\caption{%
		Products of the local density of states for the parallel and antiparallel configurations as a function of the chemical potential $E$ (measured from the Fermi energy $E_{\text{F}}$).
		(a) The case with the third and fifth layers.
		(b) The case with the second and sixth layers.
	}
	\label{fig:femgofe_ldos}
\end{figure}
\par
The products of the LDOS inside the barrier for the parallel and antiparallel configurations are shown in Fig.~\ref{fig:femgofe_ldos}.
We take the product of the LDOS at the third and fifth layers in Fig.~\ref{fig:femgofe_ldos}(b),
and those at the second and sixth layers in Fig.~\ref{fig:femgofe_ldos}(c).
For both cases, we can roughly capture the transmission properties shown in Fig.~\ref{fig:femgofe_transmission} in almost all energy ranges we focus on here.
\section{Application to antiferromagnetic tunnel junction: $\mathrm{Cr}$-doped $\mathrm{RuO_{2}}$ electrode}
\label{sec:crdoped_ruo2}
In addition to the ferromagnetic tunnel junction discussed in Sec.~\ref{sec:fe_mgo_fe},
in this section, we show that the estimation of the TMR effect with the LDOS can work also in the antiferromagnetic tunnel junctions.
Based on Ref.~\cite{Tanaka2024_PhysRevB_110_064433},
here we discuss the antiferromagnetic TMR effect using the rutile $\mathrm{RuO_{2}}$ electrode with $\mathrm{Cr}$-doping.
Recently an experiment have shown that the $\mathrm{Cr}$-doping into the altermagnetic candidate $\mathrm{RuO_{2}}$ will be useful for controlling the magnetism in the $\mathrm{RuO_{2}}$ systems~\cite{Wang2023_NatCommun_14_8240};
the anomalous Hall effect in zero magnetic field has been observed in the $\mathrm{Cr}$-doped $\mathrm{RuO_{2}}$,
which should be contrasted to the observation of the anomalous Hall component in $\mathrm{RuO_{2}}$ in a magnetic field~\cite{Feng2022_NatElectro_5_735}.
We discuss the TMR effect in the $\mathrm{Ru}_{1-x}\mathrm{Cr}_{x}\mathrm{O}_{2}(001)/\mathrm{TiO_{2}}(001)/\mathrm{Ru}_{1-x}\mathrm{Cr}_{x}\mathrm{O}_{2}$ MTJ.
\subsection{System and method}
\begin{figure}[tbh]
	\centering
	\includegraphics[width=80mm]{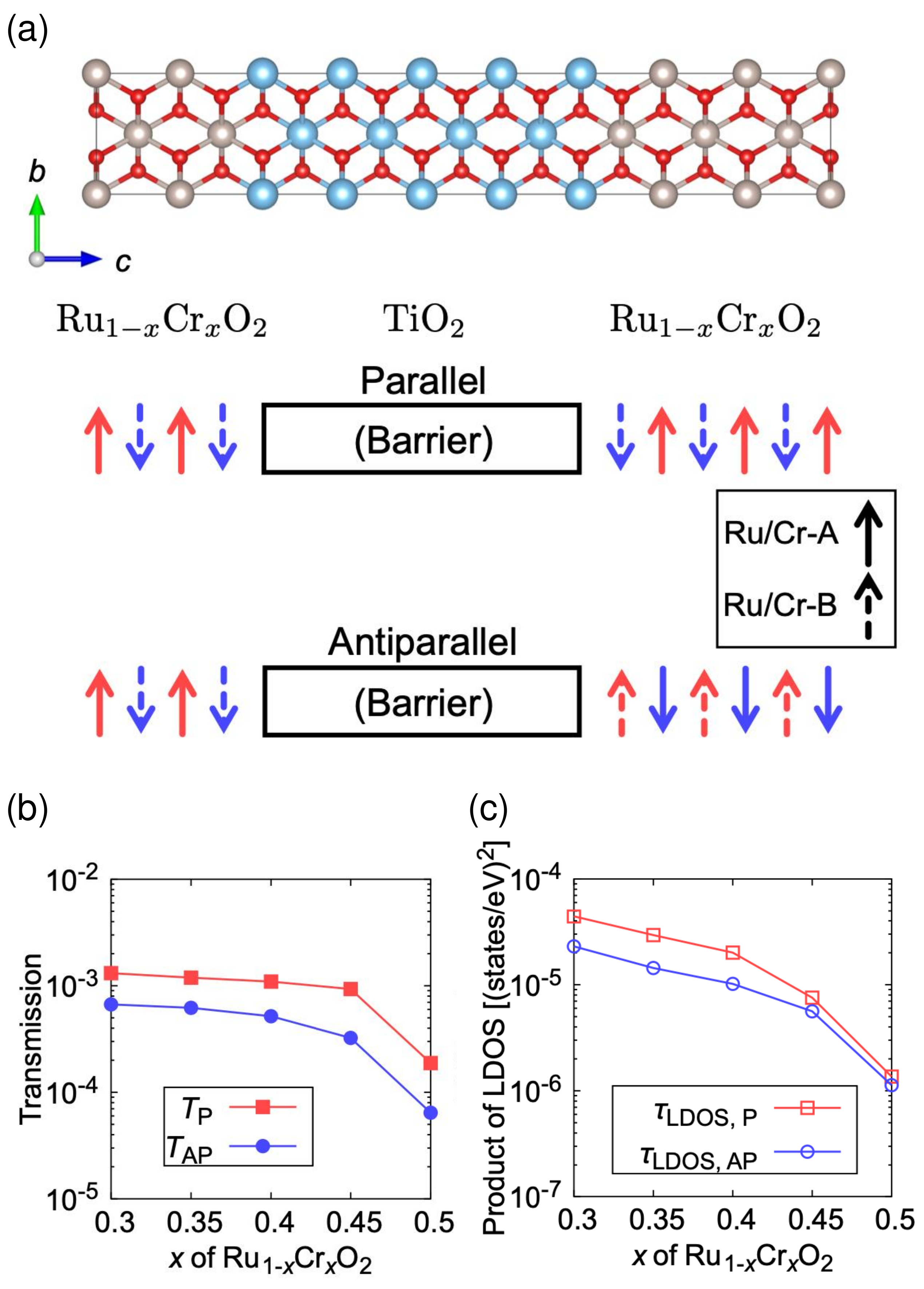}
	\caption{%
		Tunnel magnetoresistance effect in the $\mathrm{Ru}_{1-x}\mathrm{Cr}_{x}\mathrm{O}_{2}(001)/\mathrm{TiO_{2}}(001)/\mathrm{Ru}_{1-x}\mathrm{Cr}_{x}\mathrm{O}_{2}$ magnetic tunnel junction (MTJ).
		(a) Crystal structure of the scattering region used in the center of the MTJ, 
		and the schematics of the parallel and antiparallel configurations of the MTJ.
		Arrows and arrows described by broken lines are the magnetic moments of the two inequivalent Ru/Cr sites, Ru/Cr-A and Ru/Cr-B.
		(b), (c) Cr-concentration dependence of (b) the total transmission in the MTJ and (c) the product of the local density of states calculated by Eq.~(\ref{eq:ldos_product}), for the parallel and antiparallel configurations.
		Figures are adopted from Ref.~\cite{Tanaka2024_PhysRevB_110_064433} (\copyright American Physical Society (2024)).
	}
	\label{fig:rucro2_mtj}
\end{figure}

We use the $\mathrm{Cr}$-doped $\mathrm{RuO_{2}}$ as the magnetic electrode and nine MLs of the rutile-type $\mathrm{TiO_{2}}$ as the barrier.
The crystal structure of the center part of the MTJ is shown in Fig.~\ref{fig:rucro2_mtj}(a).
We take a similar manner to the case of the Fe/MgO/Fe MTJ discussed above using the QE package~\cite{Giannozzi2009_JPhysCondensMatter_21_395502,Giannozzi2017_JPhysCondensMatter_29_465901}.
Namely, we separate the MTJ into the left and right leads and the scattering region, 
perform the scf calculation for each, 
and then calculate the transmission connecting the three parts.
To calculate $\tau_{\text{LDOS}}$ (Eq.~(\ref{eq:ldos_product})), the LDOS at the fourth and sixth layers of $\mathrm{TiO_{2}}$ are used.
We note that the parallel and antiparallel configurations of the MTJ are defined by the relative orientations of the magnetic moments in the same sublattices between the left and right magnetic electrodes as shown in Fig.~\ref{fig:rucro2_mtj}(a), 
similarly to the discussion in the model calculation of the ferrimagnetic MTJ (Sec.~\ref{sec:tmr_ldos}).
For the details of the calculation, please see Ref.~\cite{Tanaka2024_PhysRevB_110_064433}.
\subsection{Results}
We show the Cr concentration dependence of the total transmission in the MTJ with parallel and antiparallel configurations, 
$T_{\text{P}}$ and $T_{\text{AP}}$, respectively, in Fig.~\ref{fig:rucro2_mtj}(b).
We confirm that $T_{\text{P}}$ takes a larger transmission than the antiparallel state in the region of $0.3 \leq x \leq 0.5$, 
which gives a positive finite TMR ratio.
The product of the LDOS with respect to the amount of Cr for the parallel/antiparallel configuration, $\tau_{\text{LDOS, P/AP}}$, 
is shown in Fig.~\ref{fig:rucro2_mtj}(c).
We find that $\tau_{\text{LDOS, P}}$ takes a larger value than $\tau_{\text{LDOS, AP}}$ in $0.3 \leq x \leq 0.5$,
which indicates that the product of the LDOS inside the barrier region $\tau_{\text{LDOS}}$ can capture the qualitative TMR property in the $\mathrm{Ru}_{1-x}\mathrm{Cr}_{x}\mathrm{O}_{2}(001)/\mathrm{TiO_{2}}(001)/\mathrm{Ru}_{1-x}\mathrm{Cr}_{x}\mathrm{O}_{2}$ MTJ.
\par
We have shown that the TMR property is roughly captured by the product of the LDOS inside the barrier by first-principles calculation of the Fe/MgO/Fe ferromagnetic MTJ (Sec.~\ref{sec:fe_mgo_fe}) and the antiferromagnetic $\mathrm{Ru}_{1-x}\mathrm{Cr}_{x}\mathrm{O}_{2}(001)/\mathrm{TiO_{2}}(001)/\mathrm{Ru}_{1-x}\mathrm{Cr}_{x}\mathrm{O}_{2}$ MTJ,
in addition to the lattice model calculation discussed in Sec.~\ref{sec:tmr_ldos}.
These results expect us to apply this method using the LDOS to computational search for the MTJs.
We may be able to avoid first-principles calculations of the transmission itself which often need huge numerical costs,
and instead we can grasp the TMR property only by calculating the LDOS of the MTJs.
\section{Summary and perspective}
In summary, we have reviewed the tunnel magnetoresistance (TMR) effect with ferrimagnetic and antiferromagnetic electrodes as well as conventional ferromagnetic electrodes.
The antiferromagnets breaking the time-reversal symmetry macroscopically can generate a finite spin splitting in the momentum space with the net spin polarization vanished,
and the momentum dependent spin splitting contributes to the realization of a finite TMR effect.
We have introduced examples of the antiferromagnetic tunnel junctions showing the TMR effect in experiments, 
in addition to the numerical studies.
\par
In the latter part, we have introduced a method to estimate the TMR effect using the local density of states inside the barrier,
and its demonstration with the density functional theory calculation.
This approach may contribute to further exploration of the magnetic tunnel junctions (MTJs), 
particularly the antiferromagnetic tunnel junctions which cannot be understood by the original Julliere model.
\par
Recently, many antiferromagnets with broken time-reversal symmetry have been found~\cite{Nomoto2024_PhysRevB_109_094435}.
Those antiferromagnets are expected to show ferromagnetic-like behaviors including the TMR effect;
we may have a much room for developing antiferromagnetic MTJs which is comparable or even superior to the ferromagnetic MTJs.
Furthermore, recent advances in antiferromagnetic spintronics have shown that the antiferromagnetic order can be manipulated by various external stimuli such as the electric current~\cite{Zelezny2014_PhysRevLett_113_157201,Liu2018_NatElectro_1_172,Chen2018_PhysRevLett_120_207204,Meinert2018_PhysRevApplied_9_064040,Hajiri2019_ApplPhysLett_115_052403,Tsai2020_Nature_580_608,Tsai2021_SmallSci_1_2000025,Higo2022_Nature_607_474,Deng2023_NatlSciRev_10_nwac154}.
By combining the antiferromagnetic TMR effect with these capability of manipulating the antiferromagnetic order,
we may be able to develop all antiferromagnetic spintronic devices,
which can possess the merit such as a high-speed operation or a smaller stray field.
\ack
We thank Xianzhe Chen, Tomoya Higo, Shinji Miwa, and Satoru Nakatsuji for fruitful discussions and collaborative work.
KT is grateful to Takashi Koretsune and Yuta Toga for discussions.
This work was supported by JST-Mirai Program (Grant No.~JPMJMI20A1), JST-CREST (Grant No.~JPMJCR23O4), JST-ASPIRE (Grant No.~JPMJAP2317), 
JSPS-KAKENHI (Grant No.~21H04437, No.~JP21H04990, No.~JP22H00290, No.~JP24K00581),
and the RIKEN TRIP initiative (RIKEN Quantum, AGIS, Many-body Electron Systems).
We use the \textsc{VESTA}~\cite{Momma2011_JApplCryst_44_1272} software to visualize the crystal structure, 
whose input file is generated with the help of the \textsc{XCrySDen} software~\cite{Kokalj1999_JMolGraphicsModelling_17_176}.
\section*{References}
\bibliographystyle{iopart-num}
\providecommand{\newblock}{}

\end{document}